\documentstyle[11pt]{article}
\newcounter{bla}

\def\ni{\noindent }
\def\eq #1{Eq.(\ref{#1})}       
\def\l{\left}                   
\def\r{\right}                  
\def\fr{\frac}                  

\def\se #1{sec. \ref{#1}}

\def\st{$^{\rm st\,}$}           
          
\def\fz{f_0}
\def\fo{f_1}
\def\ft{f_2}
\def\f3{f_3}

\def\flz{\tilde{\fz}}
\def\flo{\tilde{\fo}}
\def\flt{\tilde{\ft}}
\def\fl3{\tilde{\f3}}

\def\s3t{\tilde{{s_3}}}

\def\FPQ{$\{F,\,P,\,Q\}$}

\def\C{{\cal C}}

\def\ODEtools{{\it ODEtools}}

\def\e{{\rm e}}                  

\def\y1{\mbox{$y'$}}             

\def\|{\'\i}
\begin{document}
%

\title{Abel ODEs: Equivalence and Integrable Classes}

\author{E.S. Cheb-Terrab$^{a,b,c}$ and A.D. Roche$^d$ }

\date{}
\maketitle
\thispagestyle{empty}


\centerline {\it $^a$Department of Mathematics}
\centerline {\it University of British Columbia, Vancouver, Canada.}

\medskip
\centerline {\it $^b$CECM, Department of Mathematics and Statistics}
\centerline {\it Simon Fraser University, Vancouver, Canada.}

\medskip
\centerline {\it $^c$Department of Theoretical Physics}
\centerline {\it State University of Rio de Janeiro, Brazil.}

\medskip
\centerline {\it $^d$Symbolic Computation Group}
\centerline {\it Computer Science Department, Faculty of Mathematics}
\centerline {\it University of Waterloo, Ontario, Canada.}

\bigskip
\centerline{Original version: 17 July 1999. Revised version: 5 January 2000}

\begin{abstract}

A classification, according to invariant theory, of non-constant invariant
Abel ODEs known as solvable and found in the literature is presented. A set
of new integrable classes depending on one or no parameters, derived from
the analysis of the works by Abel, Liouville and Appell
\cite{abel,liouville3,appell}, is also shown. Computer algebra routines were
developed to solve ODEs members of these classes by solving their related
equivalence problem. The resulting library permits a systematic solving of
Abel type ODEs in the Maple symbolic computing environment.


\end{abstract}






\bigskip
\hspace{1pc}
{\bf PROGRAM SUMMARY}
\bigskip

\begin{small} \noindent {\em Title of the software package: Extension to the
Maple \ODEtools\ package}\\[10pt]
{\em Catalogue number:} (supplied by Elsevier)\\[10pt]
{\em Software obtainable from:} CPC Program Library, Queen's University of
Belfast, N. Ireland (see application form in this issue)\\[10pt]
{\em Licensing provisions:} none\\[10pt]
{\em Operating systems under which the program has been tested:} UNIX,
Macintosh, Windows (95/98/NT).\\[10pt]
{\em Programming language used:} {\ Maple V} Release 5\\[10pt]
{\em Memory required to execute with typical data:} 16 Megabytes.\\[10pt]
{\em Nature of mathematical problem}\\ Analytical solving of Abel type first order
ODEs having {\it non-constant} invariant.\\[10pt]
{\em Methods of solution}\\ Solving the equivalence problem between a given
ODE and representatives of a set of non-constant invariant Abel ODE classes
for which solutions are available.\\[10pt]
{\em Restrictions concerning the complexity of the problem}\\ The
computational routines presented work when the input ODE belongs to one of
the Abel classes considered in this work. This set of Abel classes can be
extended, but there are classes - depending intrinsically on many parameters -
for which the solution of the equivalence problem, as presented here, may lead to large and
therefore untractable expressions. When the invariants of a given
Abel ODE depend on analytic functions, the success of the routines
depends on Maple's ability to normalize these invariants and
recognize zeros (this is well implemented in Maple, but it may
nevertheless not work as expected in some cases). Also, when the solution for the
class parameter depends on other algebraic symbols entering
the ODE being solved, the routines can determine this dependency only when it has
rational form.
\\[10pt]
{\em Typical running time}\\ The methods being presented here have been implemented 
in the framework of the \ODEtools\ Maple package. On the average, over 
Kamke's \cite{kamke} first order Abel examples (see \se{performance}), 
the ODE-solver of \ODEtools\ is now spending $\approx 6$ sec. per ODE when 
{\it successful}, and $\approx 11$ sec. when {\it unsuccessful}.
The timings in this paper were obtained using Maple R5 on a Pentium II 400 - 
128 Mb. of RAM - running Windows98.\\[10pt]
{\em Unusual features of the program}\\ These computational routines are
able - in principle - to integrate the infinitely many members of all the
{\it non-constant} invariant Abel ODE classes considered in this work. Concretely, when a
given Abel ODE belongs to one of these classes, the routines
can determine this fact, by solving the related equivalence problem,
and
then use that information to return a closed form solution without requiring further
participation from the user. The ODE families that are covered include, as
particular cases, {\it all} the Abel solvable cases presented in Kamke's
and Murphy's books, as well as the Abel ODEs member of other classes not
previously presented in the literature to the best of our knowledge. After
incorporating the new routines, the ODE solver of the ODEtools package
succeeds in solving 97 \% of Kamke's first order examples. \end{small}

\newpage

\hspace{1pc}
{\large {\bf LONG WRITE-UP}}

\section{Introduction}

From some point of view, after Riccati type ODEs, the simplest first order
ordinary differential equations (ODEs) are those having as right hand side
(RHS) a third degree polynomial in the dependent variable, also called {\it
Abel type} ODEs\footnote{For convenience, in this work, by ``Abel ODEs" we
mean Abel ODEs of first kind,  since Abel ODEs of second kind can always be
transformed into first kind by a simple change of variables.}

\begin{equation}
\y1 = \f3\,  y^3 + \ft\,  y^2 + \fo\,  y + \fz
\label{abel}
\end{equation}

\ni where $y\equiv y(x)$, and $\fz$, $\fo$, $\ft$ and $\f3$ are analytic
functions of $x$. As opposed to Riccati ODEs, for which integration
strategies can be built around their equivalence to second order linear
ODEs, Abel ODEs admit just a few available integration strategies,
most of them based on the pioneering works by Abel, Liouville and Appell
around 100 years ago \cite{abel,liouville3,appell}. In those works
it was shown that Abel ODEs can be organized into equivalence classes. Two Abel
ODEs are defined to be equivalent if one can be obtained from
the other through the transformation

\begin{equation}
\{x = F(t),\ \ \ \ y(x)= P(t)\,  u(t) + Q(t)\}
\label{tr}
\end{equation}

\ni where $t$ and $u(t)$ are respectively the new independent and dependent
variables, and $F$, $P$ and $Q$ are arbitrary functions of $t$ satisfying
$F'\neq0$ and $P\neq 0$.

Integration strategies were then discussed in \cite{liouville3,appell},
around objects {\it invariant} under \eq{tr}\footnote{The invariants change
in form for $F(t) \neq t$, but keep their value. See \eq{I}.} (herein called the 
{\it invariants}) which can be built with the coefficients $\{ \f3,\, \ft,\,
\fo,\, \fz \}$ and their derivatives. To each class there corresponds a
different set of values of these invariants, and actually any one of them
(we shall pick one and call it the {\it invariant}) is enough to
characterize a class. A simple integrable case happens when the invariant
{\it is constant}\footnote{There exists one invariant such that if it
is constant then the other invariants are as well.}; the solution to the ODE then
follows straightforwardly in terms of quadratures, as explained in textbooks
\cite{kamke,murphy}. On the contrary, when the invariant is not constant,
just a few integrable cases are known and the formulation of solving
strategies based on the equivalence between two such Abel ODEs, one of which
is integrable, appears to be only partially explored in the literature, and
not explored in general in computer algebra systems.

Having this in mind, this paper concerns Abel ODEs with non-constant
invariant and presents:

\begin{enumerate}

\item A discussion and classification of the integrable Abel ODEs found both in
Kamke's book and in the works from the late 19$^{th}$ and early 20$^{th}$ century by Abel, Appell,
Liouville and other sources; \label{ii_2}

\item A set of new integrable Abel ODE classes - some depending on
arbitrary parameters - derived from those aforementioned works; \label{ii_5}

\item An explicit method of verifying or denying the equivalence between two
given Abel ODEs, one of which we know how to solve since it represents one of
the above mentioned classes; and in the positive case, a way to determine the
function parameters $F$, $P$ and $Q$ of the transformation \eq{tr} which maps
one into the other; \label{ii_0}

\item A computational scheme to resolve the equivalence problem in
the case of {\it parameterized} classes, including the determination of
the value for the class parameter when the equivalence exists;\label{ii_6}

\item A set of computer algebra (Maple) routines implementing
the algorithms presented in items (\ref{ii_0}) and (\ref{ii_6}) above, to
systematically solve - in principle - any Abel ODE belonging to one of the classes,
parameterized or not, presented here and for which a closed form solution
is known (items (\ref{ii_2}) and (\ref{ii_5}) above).\label{ii_1}


\end{enumerate}

\ni Item ({\ref{ii_2}) is interesting since the Abel ODEs shown in
textbooks in general, including Kamke's book, are displayed without
further classification, and in fact many of them belong to the same class.
This classification in terms of invariant theory is also necessary in a
computational scheme for solving Abel ODEs as the one being presented, and
we have not found it in other references. The integrable classes mentioned in
({\ref{ii_5}) are new to the best of our knowledge, although directly or
indirectly derived from previous works. The formulation of the equivalence
problem mentioned in (\ref{ii_0}) is the one given by Liouville in
\cite{liouville3}, is systematic and does not involve solving any
auxiliary differential equations\footnote{An approach somewhat similar to
this one by Liouville is discussed in \cite{schwarz}.}. Concerning item
(\ref{ii_6}), the idea can be viewed as a way of avoiding the untractable
expressions which one would encounter when making direct use of
Liouville's strategy with {\it parameterized} classes. The strategy
presented is applicable when there exists a solution for some numerical
values of the parameter, or when this parameter is a rational function of
other symbols entering the input ODE. Regarding
item (\ref{ii_1}), the implementation presented here is, as far as we
know, unique in computer algebra systems in its ability to solve
non-constant invariant, parameterized or not, Abel ODE classes.


The paper is organized as follows. In \se{theory}, the basic definitions
and the classic formulation of the equivalence problem in terms of
invariants is reviewed and shown to
apply straightforwardly to the case of a non-parameterized class.
In \se{parameterized_seed_method} it is shown how
these ideas can be complemented by taking advantage of computers to tackle
the equivalence problem in the case of a parameterized class. Section
\ref{solvable_seeds} presents a classification of the integrable classes
we have found in the literature with some additional comments as to their
derivation. In \se{new_seeds} new integrable Abel classes are presented.
In \se{performance}, a test-suite for the routines presented is discussed
and statistics are shown describing the performance with this test-suite
as well as with Kamke's first order examples. Finally, the conclusions
contain some general remarks about this work and its possible extensions.

Additionally, we present in the Appendix a table listing the distinct Abel ODE
classes that we have found, representative ODEs from each class, and their
respective solutions.


\section{Classical Theory for Abel ODEs}
\label{theory}


In general, Abel type ODEs can be studied
using two related concepts: {\it invariants} and ODE {\it equivalence classes}.
We define two Abel ODEs to be equivalent\footnote{For a more formal
definition of {\it class} see \cite{olver}} if 
one can be obtained from the other using a transformation of
the form \eq{tr}. The equivalence class containing a given ODE is then the
set of all the ODEs equivalent to the given one.
We note that although the infinitely many members of a class can be mapped
into each other by using \eq{tr}, there are also infinitely many disjoint
Abel classes (\eq{tr} is not sufficient to map {\it any} Abel ODE into a given one).

To each class one can associate an infinite sequence of absolute
invariants \cite{liouville3,appell}. To see this, consider two Abel ODEs, the first \eq{abel}, the
second obtained from \eq{abel} through the transformation \eq{tr}

\begin{equation}
u' = \fl3\,  u^3 + \flt\,  u^2 + \flo\,  u + \flz
\label{abel2}
\end{equation}

\ni where the coefficients $\{\flz,\, \flo,\, \flt,\, \fl3 \}$, are related
to the those of \eq{abel} by

\begin{eqnarray}
{\flz} & = & {\frac { F^{'}\l( {\vrule height .7em width0em depth .54 em}
    \fz(F) +\fo(F) \,Q+\ft(F) \,{Q}^{2} +\f3(F) \,{Q}^{3} \r) - Q^{'}}{P}} 
\nonumber \\*[.1in]
{\flo} & = & {\frac {P^{'}}{P}} - F^{'}\l( \fo(F) + 2\,\ft(F)\,Q+3\,\f3(F) \,{Q}^{2} \r) 
\nonumber \\*[.1in]
{\flt} & =  & P\, F^{'} \l( {\vrule height .7em width0em depth .54 em}  \ft(F)+3\,\f3(F)Q \r)
\nonumber \\*[.1in]
{\fl3} & = & {P}^{2}\,F^{'} \f3(F) 
\label{f_tilde}
\end{eqnarray}

\ni Following \cite{appell}, we call an {\it absolute invariant} of
\eq{abel} a function $I(f,x)$ of the coefficients $\{\fz,\, \fo,\, \ft,\,
\f3 \}$ and their derivatives with respect to $x$ such that, for all
$\{F,\,P,\,Q\}$ in \eq{tr},

\begin{equation}
I(\tilde{f},t) \vert _{\tilde{f}=\tilde{f}(f,t)} = I(f,x)\vert _{x=F(t)}
\label{I}
\end{equation}

\ni where ${\tilde{f}=\tilde{f}(f,t)}$ represents the coefficients
$\{\flz$, $\flo$, $\flt$, $\fl3\}$ and their derivatives with respect to
$t$, expressed using \eq{f_tilde}.

Similarly, we call a {\it relative invariant} a function, say $s$, of the coefficients
of \eq{abel} and their derivatives such that when changing variables using
\eq{tr}, the resulting expression is equal to the original one up to a
factor, say $\varphi_s$, dependent uniquely on the functions $F$, $P$, and
$Q$ in \eq{tr} and independent of the coefficients themselves
\cite{olver}:

\begin{equation}
s(\tilde{f}) \vert _{\tilde{f}=\tilde{f}(f,t)}
 = \varphi_s(F,P,Q)\ s(f) \vert _{x=F(t)}
\label{denote_s}
\end{equation}

\ni Liouville showed that in the case of Abel equations there is a relative
invariant of weight 3

\begin{equation}
s_3
\equiv \fz \f3^2
+ \frac{1}{3} \l(\frac {2\,\ft^3}{9} - \fo \ft \f3 +
\f3 \ft'- \ft \f3' \r)
\label{s3}
\end{equation}

\ni which can be used to recursively generate an infinite sequence of
relative invariants $s_{2m+1}$ of weights $2m+1$ respectively\footnote{In
the case of $s_3$, $\varphi_{s_3} = {(F' P)}^3$; the weight $n$ refers to the degree
of $\varphi_{s_n}$ with respect to ${(F' P)}$.}, through the formula

\begin{equation}
s_{2m+1} \equiv {\f3}\,{s'_{{2\,m-1}}}-\left (2\,m-1\right )s_{{2\,m-1}}\left ({
\f3^{'}}+{\fo}\,{\f3}-\fr{  {{\ft}}^{2} }{3}\right )
\label{s2m+1}
\end{equation}

\ni As is clear from this definition, the product of two relative invariants
respectively of weights $n$ and $m$ is a relative invariant of weight $n+m$,
and by dividing any two relative invariants of equal weight one can generate
an infinite sequence of absolute invariants

\begin{equation}
I_1= \frac {s_5^3}{s_3^5},\
I_2= \frac {s_7 s_3}{s_5^2},\
I_3= \frac {s_9 }{s_3^3},\
\mbox{etc...}
\label{invariants}
\end{equation}

\ni In \cite{appell}, Appell showed that this sequence can also be obtained
from two basic absolute invariants - say $J_1,\, J_0$, by expressing $J_1$
as a function of $J_0$ and then differentiating the result with respect to
$J_0$. As a consequence, if $I_1$
is constant then all the other ones will be too. This fact allows
one to identify the constant character of the invariants in \eq{invariants}
by looking at just the first one. We note also
that there are infinitely many {\it different} classes having $I_1$
constant, related to the infinitely many possible constant values $I_1$ can
have.

\subsection{Integration strategy}

A description of a method of integration when the invariants are
constant\footnote{In \cite{schwarz} it is also shown that in the constant
invariant case the problem can also be formulated in terms of the
symmetries of these ODEs.} is found in the works by Abel \cite{abel},
Liouville \cite{liouville3} and Appell \cite{appell}. In such a case, {\it
all} members of the class can be systematically mapped into a {\it
separable} first order ODE (representative of the class), by appropriately
choosing $F$, $P$ and $Q$ in \eq{tr}; see for instance \cite{kamke} and
\cite{murphy}.

A quite different situation happens when $I_1$ is not constant. In such a
case, relatively few integrable Abel ODEs are known, and the integration methods
used to solve each of them depend in an essential way on non-invariant
properties of the coefficients $f$. Those methods are then useless for
solving the other infinitely many members of the same classes, unless one
can solve the related equivalence problems; i.e., determining - when they
exist -the values of $F$, $P$ and $Q$ in \eq{tr} linking two Abel ODE which
belong to the same class.

\subsection{Identifying an ODE as member of a given Abel ODE class}
\label{equivalence}

Consider two Abel ODEs; the first one given by \eq{abel}, and a second one
being of the same form, but with coefficients $\flz$, $\flo$, $\flt$ and $\fl3$. The
problem now is to determine whether the second Abel ODE can be obtained from
\eq{abel} by changing variables using \eq{tr}.
This problem can be formulated by equating the coefficients between the {\it
transformed} equation, obtained by applying the transformation \eq{tr} to
\eq{abel}, and the second Abel ODE, resulting in \eq{f_tilde}, which can be
seen as an ODE system for $\{F,\,P,\, Q\}$. To solve this system, following
Liouville \cite{liouville3}, we first note that the {\it absolute}
invariants corresponding to the two Abel ODEs don't depend on $P$ or $Q$
(see previous section). Hence the function $F$ entering \eq{tr} can be
obtained by just running an elimination process using two of these absolute
invariants, for instance $I_1={s_5^3}/{s_3^5}$ and $I_2=s_5
s_7/s_3^{4}$:

\begin{equation}
0
 = 
\displaystyle
\frac {\tilde{s_5}^3}{\tilde{s_3}^5} - \frac {s_5^3}{s_3^5}\vert _{x=F(t)}
\ \ \ \ \ \ 
0
 = 
\displaystyle
\frac {\tilde{s_5}\tilde{s_7}}{\tilde{s_3}^4} - \frac {s_5 s_7}{s_3^4}\vert _{x=F(t)}
\label{sysAB}
\end{equation}

\ni As discussed in \cite{liouville3,appell}, the {\it existence} of a
common solution $F(t)$ to both equations above (such that $F'\neq 0$) is
the necessary and sufficient condition for the existence of a
transformation \eq{tr} relating the two Abel ODEs. Once $F$ is known, the
system \eq{f_tilde} becomes trivial in that $P$ and $Q$ can be
re-expressed in terms of $F$ by performing fairly simple calculations. In
the case of interest of this work - non-constant invariant\footnote{When
the invariant is non-constant, $s_3 \neq 0$.} - the resulting expressions
are:

\begin{equation} P(t)={\frac {{\it
F'}\,{\fl3}^{2}s_{{3}}}{{f_{{3}}}^{2}\s3t}} \vert _{x=F(t)} \ \ \ \ \ \ \ \
Q(t)={\frac {{\it F'} \flt\,\fl3 s_{{3}}-f_{{2}}f_{{3}}\s3t}
{3\,{f_{{3}}}^{2} \s3t}} \vert _{x=F(t)}
\label{ans_PQ}
\end{equation}

\ni where $\{f_i,\, \tilde{f}_i\}$ with $i:0 \rightarrow 3$ are the
coefficients of the two Abel equations, $s_3$ is the relative invariant
\eq{s3} expressed in terms of $f_i$ and $\s3t = s_3\vert_{f_i=\tilde{f_i}}$.

Concerning the explicit solution $F(t)$ for \eq{sysAB}, we note that our
interest in solving the equivalence problem is in that it leads directly to
the solution of other members of an Abel class, when the solution to a
representative of the class is known. In turn, all the solvable classes we
are aware of have a representative with rational coefficients (see the Appendix), and hence also
rational invariants\footnote{On the other hand, there is no reason to expect
that the second of the two Abel ODEs being tested for equivalence also has
rational invariants. If however the invariants of {\it both} Abel ODEs are rational in $t$
{\it and} the coefficients are {\it numbers}, then it is also possible to determine
$F(t)$ by performing a rational function decomposition as mentioned in
\cite{schwarz}.}. Hence, assuming that one of the two Abel ODEs has rational
coefficients and that \eq{sysAB} was obtained by applying \eq{tr} to it, the
system \eq{sysAB} will always be polynomial in $F(t)$. In such a case, when
a common solution $F(t)$ to both equations exists, the resultant between
these polynomials will be zero \cite{liouville3}; i.e.: there will be a
common factor, depending on $F$ and $t$ and representing the common solution,
which can be obtained by calculating the greatest common divisor (GCD)
between the two equations in \eq{sysAB}. Conversely, if that GCD does not
depend on $F$, a transformation \eq{tr} linking the input equation to
\eq{abel} does not exist. That the dependence on $F$ of this GCD is a
necessary condition for the existence of the desired transformation \eq{tr}
is a consequence of the validity of \eq{I} and hence the system
(\ref{sysAB}). A proof of its sufficiency was given by Appell in
\cite{appell}.

The whole process just described to determine the equivalence between two
given Abel ODEs, one of which is rational in $x$, can be summarized as
follows:

\begin{enumerate}


\item Calculate two absolute invariants, set up the system \eq{sysAB}, and
calculate the GCD between the two equations;\label{set_system}

\item When this GCD does not depend on $F$, the ODEs don't belong to the same
class; otherwise determine an explicit expression for $F(t)$ from the result
of the GCD calculation;\label{determine_F}

\item Plug this value for $F$ into the formulas \eq{ans_PQ} to determine the
values of $P(t)$ and $Q(t)$, arriving in this way at the transformation
\eq{tr} mapping one Abel ODE into the other.

\end{enumerate}

\medskip
\ni {\it Example:}
\medskip

Consider the two non-constant invariant Abel ODEs
\begin{equation}
%
%
\y1=-\fr{1}{2(x+4)}\,\l(xy^3+ y^2\r)
\label{ex1_abel1}
\end{equation}

\begin{equation}
%
%
%
\y1=
{\frac { \left ({\it f'}\,x-f\right )} {2\,(f+3\,x)}}
\l(\left (x-f\right ) {y}^{3} + {y}^{2}\r)
-{\frac {y}{x}}
\label{ex1_abel2}
\end{equation}

\ni where in the above $f\equiv f(x)$ is an analytic (arbitrary) function. As in
the typical situation one of these ODEs has invariants rational in $x$ and we know its
solution; i.e. for \eq{ex1_abel1} we have

\begin{equation}
C_1+{\frac {\sqrt {{y}^{2}x-4\,y-1}}{y}}+2\,\arctan\l({\frac {1+2\,y}{
\sqrt {{y}^{2}x-4\,y-1}}}\r)=0
\label{ans_ex1_abel1}
\end{equation}

\ni where $C_1$ is an arbitrary constant. We would then like to determine
whether there are functions \FPQ\ so that \eq{ex1_abel1} transforms
under \eq{tr} into \eq{ex1_abel2}, and if so, determine \FPQ\ and use
them together with \eq{ans_ex1_abel1} to build the answer to \eq{ex1_abel2}.
For this purpose, we start (step (\ref{set_system})) by computing the
relative invariants $s_3,\, s_5,\, s_7$, leading to \eq{sysAB}

\begin{equation}
\begin{array}{lcl}
0
& = &
{\frac {\left (87\,ft+9\,{f}^{2}+184\,{t}^{2}\right )^{3}}{t\left (9\,
f+31\,t\right )^{5}}}-{\frac {\left (280+105\,F+9\,{F}^{2}\right )^{3}
}{\left (40+9\,F\right )^{5}}}
\\*[.13in]
0
& = &
{\frac {\left (81\,{f}^{3}+1431\,{f}^{2}t+7185\,f{t}^{2}+10903\,{t}^{3
}\right )\left (9\,f+31\,t\right )}{\left (87\,ft+9\,{f}^{2}+184\,{t}^
{2}\right )^{2}}}-{\frac {\left (1674\,{F}^{2}+10290\,F+81\,{F}^{3}+
19600\right )\left (40+9\,F\right )}{\left (280+105\,F+9\,{F}^{2}
\right )^{2}}}
\end{array}
\end{equation}

\ni where in the above $F\equiv F(t)$ is the function we are looking for
and $f$ is taken at $x=t$. We recall that when \eq{ex1_abel2} has
non-constant invariant the denominators in above are not zero since $s_3
\neq 0$. Calculating the GCD between the numerators of the expressions
above (step (\ref{determine_F})) and equating this GCD to zero, we
obtain\footnote{We note that when $f(t)$ is an algebraic mapping involving
varied analytic functions, Maple's procedures to simplify and put them in
normal form may fail and consequently the GCD computation may not be
successful.}

\begin{equation}
27\,(t+tF-f) = 0
\end{equation}

\ni from where the common solution $F(t)$ to both equations is given by
\begin{equation}
F(t)={\frac {f(t)}{t}}-1
\end{equation}

\ni Substituting this value of $F$ into \eq{ans_PQ}, a transformation of the
form \eq{tr} mapping \eq{ex1_abel1} into \eq{ex1_abel2} is finally given by

\begin{equation}
\{x = {\frac {f(t)}{t}}-1,\ \ \ \ y(x)= t\, u(t)\}
\end{equation}

\ni from where by changing variables in \eq{ans_ex1_abel1} using the
transformation above and renaming the variables ($t\rightarrow x,\;
u\rightarrow y$), the solution to \eq{ex1_abel2} is obtained

\begin{eqnarray}
C_1
+ {\fr{\sqrt {\left ({\frac {f}{x}}-1\right ){x}^{2}{y}^{2}-4\,xy-1}}
{xy}}
+ 2\,\arctan\l({\frac {\left(1+2\,xy\right)}
{\sqrt {\left ({\frac {f}{x}}-1\right ){x}^{2}{y}^{2}-4\,xy-1}}}\r)
= 0
\end{eqnarray}

\section{Parameterized Abel ODE classes}
\label{parameterized_seed_method}


We formulate here the equivalence problem in the case of {\it
parameterized} classes. By ``parameterized class" we mean an (Abel) ODE
class depending on symbolic parameters which {\it cannot be removed} by
changing variables using \eq{tr}. The interest in {\it parameterized}
solvable classes is clear: to each set of values of the parameters
corresponds - roughly speaking - a different Abel class\footnote{There may
be particular different sets of values for which the resulting ODEs will
nevertheless belong to the same class.}. Hence, a formulation of the
equivalence problem for parameterized classes enables one to solve all the
members of infinitely many classes at once.

In order to simplify the discussion, we consider the problem of an Abel
ODE class depending on just one parameter\footnote{The integrable classes
presented in the literature depend at most on one parameter (see
\se{solvable_seeds}).}, say $\C$. Also, we distinguish between two
different types of problems: one is when the equivalence problem has a
solution for a specific {\it numerical} value of $\C$; the other happens
when to have a solution it is required that $\C$ assumes {\it symbolic}
values, for instance in terms of other {\it symbols} entering the input
ODE. We discuss first the {\it numerical} case, and in the next subsection
we show how the {\it symbolic} case can be mapped into many {\it
numerical} problems - when the parameter depends on other symbols in a
{\it rational} manner - by using rational interpolation methods.

\subsection{Solution for some {\it numerical} value of $\C$}
\label{numerical_C}

To facilitate the exposition we present the discussion around a concrete
example. Consider the equivalence problem between a given Abel ODE,
for instance,

%
\begin{equation}
\y1=8\,{\frac {\left (1-{x}^{4}-{x}^{8}\right ){y}^{3}}{{x}^{7}}}+4
\,{\frac {{y}^{2}}{{x}^{4}}}+{\frac {y}{x}}
\label{any}
\end{equation}

\ni and the one presented in Abel's memoires \cite{abel}

\begin{equation}
\y1={\frac {\left (\C \,{x}^{4}+{x}^{2}+1\right ){y}^{3}}{{x}^{3}}}
+{y}^{2}
\label{abel1}
\end{equation}

\ni If this equivalence exists, then it exists for a {\it specific} value of
the parameter $\C $ since there is no solution for arbitrary $\C$ (the
existence of such a solution would mean the class does not really depend
on any parameter). Hence, the common solution $F(t)$ to the system
\eq{sysAB} will not show up until the correct value of $\C$ is
determined\footnote{There may be more than one solution $\C$.},
invalidating the itemized algorithm of the previous section.

A natural alternative to this problem would be to take one more absolute invariant,
for instance, $ s_3 s_7 / s_5^2$, so that our system \eq{sysAB} becomes


\begin{equation}
0
 = 
\displaystyle
\frac {\tilde{s_5}^3}{\tilde{s_3}^5} - \frac {s_5^3}{s_3^5}\vert _{x=F(t)}
\ \ \ \ 
0
 = 
\displaystyle
\frac {\tilde{s_3}\tilde{s_7}}{\tilde{s_5}^2} - \frac {s_3 s_7}{s_5^2}\vert _{x=F(t)}
\ \ \ \
0
 = 
\displaystyle
\frac {\tilde{s_5}\tilde{s_7}}{\tilde{s_3}^4} - \frac {s_5 s_7}{s_3^4}\vert _{x=F(t)}
\label{system}
\end{equation}

\ni and search for a solution to this problem such that $F\,' \neq 0,\ \C \,'
= 0$. For that purpose, eliminate $\C$ from the first and second
expressions above by taking the resultant with respect to $\C$, obtaining
- say - $R_1$. In the same way, eliminate $\C$ from the first and the
third expressions of \eq{system} obtaining $R_2$. Hence, when a solution
exists, the resultant between $R_1$ and $R_2$ with respect to $F$ will
vanish. In other words, the algorithm of the previous section will work if
instead of performing the calculations over the expressions \eq{sysAB} we
perform them over $R_1$ and $R_2$. The
GCD between $R_1$ and $R_2$ will then return the factor depending on both
$F$ and $t$, whose solution is the function $F(t)$ we are interested in.

This method, simple and correct in theory, unfortunately does not work in
practice because the expressions tend to grow in size so much that the
computation of the first of these three resultants may not be possible,
even with a simple example such as the one shown above. The problem
resides in the fact that multivariate GCDs and resultants are quite
expensive operations for the current symbolic computation environments.

An alternative to this problem consists of reducing it to a sequence of
bivariate GCD and resultant calculations, for which the available
algorithms are relatively fast. The idea can be summarized as follows.

\begin{enumerate}

\item \label{step1} From the previous considerations, when a solution to the
equivalence problem exists, the resultant between any two of the expressions
in \eq{system} {will not vanish for any value of $t$}, since we haven't
introduced the correct (unknown at this point) value of $\C$. Hence, if we
insert in \eq{system} a numerical value\footnote{We note there may exist
``invalid evaluation points"; roughly speaking to avoid this problem this
evaluation point must not cancel any of the coefficients of the variables remaining in
the system - see \cite{keith}.} for $t$ and calculate the GCD between
any two of the resulting expressions, this GCD {\it cannot contain any
factor depending on $F$}. This gives us a first ``existence
condition" test for the solution before proceeding further;

\item When \eq{system} evaluated at $t=\mbox{number}$ passed the test of the
previous step, take two of the resulting three expressions and calculate their
resultant with respect to $F$, obtaining, say, $\tilde{R}_1$. Then take a {\it
different} pair and calculate their resultant with respect to
$F$ again, obtaining, say, $\tilde{R}_2$. Neither of these resultants will
vanish since the GCD calculations of the previous step showed no
factor depending on $F$. Also, the calculation of $\tilde{R}_1$ and
$\tilde{R}_2$ is now quite simpler since the expressions do not involve $t$;\label{step2}

\item \label{step3} Then if a solution to the problem exists, the GCD
between $\tilde{R}_1$ and $\tilde{R}_2$ will yield a factor depending on
$\C$; equating it to zero and solving it for $\C$ will give the {\it
common solution $\C$ for $\tilde{R}_1$ and $\tilde{R}_2$}. More precisely,
what we will get in this way is a set of {\it candidates} (including among
them the correct value) for $\C$; not all of them will necessarily lead to
a solution $F(t)$ to the original problem;

\item \label{step4} We now plug these candidates for $\C$ into
\eq{system}, one at a time, receiving a system of three expressions
involving again only two unknowns, now $F$ and $t$. If there is a common
solution $F(t)$ to these expressions, the resultant with respect to F
between any two of them will vanish. Hence, the GCD between those two
expressions will contain a factor depending both on $F$ and $t$; equating
this factor to zero and solving for $F$ leads to the solution $F(t)$.

\end{enumerate}

\ni Returning to our example of determining the equivalence between \eq{any}
and \eq{abel1}, the itemized procedure just outlined runs as follows.

According to step (\ref{step1}), $t=0$ is tried first,
but it is found to be an invalid evaluation point. The next value of $t$ to try, $t=1$
turns out to be valid, so \eq{system} was evaluated at $t=1$; the GCDs between any
two of the three resulting expressions do not depend on $F$, so
this first test for the ``existence" of a solution passed.

Continuing with step (\ref{step2}), the calculation of $\tilde{R}_1$} and
$\tilde{R}_2$} is performed without problems concerning the size of the
expressions.

The GCD of step (\ref{step3}) results in the three factors: $36\,\C-5$,
$\C+1$ and $9\,\C-2$; equating them to zero and solving them for $\C$ we
arrive at three candidates for $\C$.

\ni In step (\ref{step4}), plugging each of these candidates one at a time into
\eq{system} and taking the GCD between two of the three resulting
expressions we note that $\C=5/36$ does not lead to any factor depending on
both $F$ and $t$, but $\C=-1$ leads to such factor: $F^2t^4-1$. So that for
$\C=-1$ the problem admits two solutions: $F=\pm 1/t^2$.

Finally, by introducing $F=1/t^2$ into the formulas for $P$
and $Q$ \eq{ans_PQ} we arrive at the transformation of the form \eq{tr}
mapping \eq{abel1} into \eq{any}

\begin{equation}
\left \{x=\fr{1}{t^2},y=-2\,{\frac {u(t)}{t}}\right \}
\end{equation}

\ni and hence by applying the same change of variables to the answer of
\eq{abel1} {\it and} substituting $C=-1$ we obtain the answer to \eq{any}.

A remark however is in order: if, in step (\ref{step3}) of the algorithm
just described, the numeric candidates for $\C$ involve fractional powers
of rational numbers, the Maple system may then enter not efficient
expensive computations, exhausting the system resources before
determining the expression $F(t)$ solving the problem in step
(\ref{step4}). The root of this limitation seems to be in the absence in
Maple of built-in normalization for such ``numeric radicals"\footnote{A
built-in normalization of radicals is implemented in the computer algebra
system ``Mathematica". For typical problems we tried where Maple exhausted
the system resources trying to determine the solution $F(t)$, we exported
the mathematical expressions involving radicals to Mathematica and noticed
that $F(t)$ was determined in this other computer algebra system in
reasonable time.}.

\subsection{Solution when the parameter $\C$ is some {\it rational}
function of other symbols}
\label{symbolic_C}

When $\C$ assumes {\it symbolic} values, for instance it depends on other
{\it symbols} - say $\{\alpha\}$ - entering the input ODE, if this
dependency is rational it is possible to map the determination of
$\C(\alpha)$ into a sequence of problems having for solution a {\it
numerical} value of $\C$. In turn each of these {\it numerical} problems
can be tackled using the algorithm of the previous subsection. The idea
consists of attributing numerical values to the symbols $\{\alpha\}$
entering the invariants in order to determine $\C(\alpha)$ by means of a
rational interpolation. To simplify the presentation we first discuss the
case when $\{\alpha\}$ consists of a {\it single} parameter, and then show
how to extend the algorithm to the case in which $\{\alpha\}$ consists of
{\it many} parameters by commenting on a concrete example. So when $\alpha$
consists of just one parameter this rational interpolation scheme is
summarized as follows:

\begin{enumerate}

\item \label{step_no_more_symbols} Take the system \eq{system} and
attribute a numerical value to $\alpha$ (check for possible wrong
evaluation points), so the resulting system depends on just $x$, $\C$
and $F$;

\item \label{step_quit} Enter step (\ref{step1}) in the algorithm of the
previous subsection and run all the steps:

\begin{enumerate}

\item If there is no solution for $\C$ and $F$ such that
$\C\,'=0$ and $F\,'\neq 0$ then quit the process - the input ODE does not
belong to this class, or the solution involves a non rational dependency
of $\C$ on $\{\alpha \}$.

\item\label{step_record} If however a solution for $\C$ and $F$ was
found, record the values for $\C$ and $\alpha$ - they represent a {\it
point} of a curve $\C(\alpha)$; 

\end{enumerate}

\item \label{step_test} Using the points recorded so far, interpolate $\C$
as a function of $\alpha$ and test if this interpolated value already
solves the problem for an {\it arbitrary} $\alpha$\footnote{This is
done by restoring the symbol ``$\alpha$" in the system obtained in step
(\ref{step_no_more_symbols}) and checking if the system is satisfied.}:

\begin{enumerate}

\item If so, the problem has been solved;

\item \label{step_last} Otherwise change the evaluation point of $\alpha$
and re-enter step (\ref{step_quit}) of this enumeration;

\end{enumerate}

\end{enumerate}

We note that the rational interpolation of $\C(\alpha)$ requires the
knowledge a priori of the polynomial degrees in $\alpha$ of both numerator
and denominator. That information is not available in advance, but we know
that these two degrees sum to $n- 1$, where $n$ is the number of points
being interpolated. So, when performing the test in step (\ref{step_test})
and before going to step (\ref{step_last}) we actually
test all possible different interpolations, starting with the maximum
possible degree for the numerator and finishing with the maximum possible
degree for the denominator.

Concerning the extension of this algorithm for the case when
$\{\alpha\}$ involves more than one parameter, for instance $\{a,\,b\}$,
this extension is easy and better illustrated with an example. Consider
the equivalence problem between the Abel ODEs


\begin{equation}
\y1={\frac {C\left (2\,{x}^{2}C-2\right ){y}^{3}-3\,C{y}^{2}+Cxy}
{1-{x}^{2}C}}
\ \ \mbox{and}\ \ \
\y1={\frac {a\left (2\,{x}^{2}a-2\,b\right ){y}^{3}-3\,a{y}^{2}b+abxy
}{b\left (b-{x}^{2}a\right )}}
\label{two_abels}
\end{equation}

\ni which solution is just the identity $x \rightarrow x,\, y \rightarrow
y$ , but only exists when $\C=\fr{a}{b}$ (we choose the identity without
loss of generality and so that the solving process is easy to follow). We
want to determine $\C=\fr{a}{b}$ (rational function of {\it two}
parameters) using the algorithm just described.

We start with step (\ref{step_no_more_symbols}), take the system
\eq{system} corresponding to \eq{two_abels} and attribute numerical values
for the first parameter, $a$, checking for possible wrong evaluation
points - we end up evaluating this system at $a=1$. However, the system
still depends on the second parameter, $b$, so we attribute numerical
values to $b$ too - check for possible wrong evaluation points - and hence
end up evaluating the system altogether at $a=1,\, b=2$. The resulting
system now only depends on $x$, $F$ and $\C$.

So we enter step (\ref{step_quit}) (actually the whole algorithm of
\se{numerical_C}) with this evaluated system of the previous step and
detect that $\C=1/2$ leads to the solution $F=x$; so a solution $\C\,'=0$
and $F\,' \neq 0$ exists.

Hence, in step (\ref{step_test}) we interpolate $\C(b)$ (so far we have just one
point, so $\C(b)=1/2$) and test this value for arbitrary $b$, verifying that
the interpolation is still incomplete: the system \eq{system} with $a=1$
and {\it arbitrary} $b$ is not satisfied by $\C=1/2,\,F=x$.

We are then in step (\ref{step_test}.\ref{step_last}); so attribute a new
numerical value to $b$, hence evaluate the system \eq{system} at
$a=1,\,b=3$ and re-enter step (\ref{step_quit}) finding that $\C = 1/3$
leads to $F = x$.

We record this new point (step (\ref{step_quit}.\ref{step_record})) and
interpolate - now using the two points obtained so far - our first
interpolation is of degree 1 in $b$: $\C(b)= (5-b)/6$. This interpolation
however does not solve the problem for arbitrary $b$. So we increase by
one the degree in $b$ of the denominator in the interpolation, resulting in
$\C(b)=1/b$, and verify that this second interpolation indeed solves the
system for arbitrary $b$ - so the interpolation for $b$ at $a=1$ is
complete.

We return then to the interpolation of $\C(a,b)$ with respect to $a$, record
that for $a=1$ there is a solution $\C(b)=1/b$, and test this solution for
{\it arbitrary} $a$ (step (\ref{step_test}) with respect to $a$), verifying
that the interpolation for $a$ is still incomplete.

Hence we re-enter step (\ref{step_no_more_symbols}) with new evaluation points
$a=2,\,b=4$, and re-start the process of determining $\C(b)$.

Following the same steps just described, for $a=2$ we find that
$\C(b)=2/b$ solves the problem for arbitrary $b$, leading to the second
point in the interpolation of $\C(a,b)$ with respect to $a$.

So we are now in step (\ref{step_test}) again, and interpolating $\C$
using the two points obtained so far we find $\C(a,b)=a/b$ - this value of
$\C(a,b)$ is verified to satisfy the system \eq{system} for arbitrary $a$,
and thus the problem has been solved.

The algorithm just described, though expensive in computations, is
successful in solving the equivalence problem in reasonable time for
typical situations (see \se{performance}). In this example, for instance,
it took 12 seconds to: setup the invariants and the system
\eq{system}, simplify this system to a normal form, run all the items of
the algorithm just described to determine $\C(a,b)$ and $F(t)$, then
determine $P(t)$ and $Q(t)$ according to \eq{ans_PQ}, and finally return
a solution to the second of the ODEs in \eq{two_abels}.


\subsubsection{Remark on the existence of multiple solutions for the class
parameter $\C(\alpha)$}

The interpolation algorithm just described is valid provided there is only
one solution curve $\C(\alpha)$; otherwise we may end up trying to
interpolate $\C(\alpha)$ using points which belong to different solution
curves, leading nowhere. In turn, the existence of many curves
$\C(\alpha)$ solving a given problem is related to the existence of
symmetries in the invariants (of the ODE representative of the class we
want to match) entering \eq{system}. Concretely, if the mapping $\{ \C
\rightarrow \kappa(\C),\ x \rightarrow \phi(x,\C)\}$ is a symmetry of
these invariants, then if $\{\C,\,F(x)\}$ leads to a solution for the
equivalence problem, consequently $\{\kappa(\C),\, \phi(F(x),\C)\, \}$ will
also lead to a (different) solution. A concrete example of this situation
is discussed in \se{simpler_invariants}.

Concerning detecting this situation, we note that if the
mapping $\{ \C \rightarrow \kappa(\C),\ x \rightarrow \phi(x,\C)\}$ is a
symmetry of the invariants, then the {\it inverse} mapping is also a
symmetry, and since these invariants are {\it rational} in both $x$ and
$\C$, the form of such a symmetry mapping is

\begin{equation}
\kappa(\C) = \fr{a\,\C+b}{c\,\C+d},\ \ \
\phi(x,\C) = \fr{f(\C)\, x +g(\C)}{h(\C)\, x +j(\C)}
\end{equation}

\ni that is, a fractional linear mapping, where {\it \{a, b, c, d\}} and
$\{f(\C),\, g(\C),\, h(\C),\, j(\C)\}$ are in principle constants and
functions to be determined. Regarding the use of interpolation methods,
the problem happens when $\kappa(\C) \neq \C$. A first manner of detecting
this problem  then consists of setting up the system of equations and
inequations

\begin{equation}
I(x,\C) = I(x,\C)_{\left \vert^{x=\phi(x,\C)}_{\C=\kappa(\C)} \right.},
\ \ 
\phi=\fr{F}{G},
\ \ 
F_{xx} = 0, \ \ G_{xx} = 0,
\ \ 
\kappa(\C) \neq \C
\label{manner_1}
\end{equation}

\ni where $I(x,\C)=s_5^3/s_3^5$ is the first invariant and $F$, $G$ and
$\kappa$ are the unknowns to be determined, and seeing if this system is
consistent. This check for consistency can be performed by simplifying
this system with respect to its integrability conditions - for this
purpose we used the {\it diffalg} \cite{boulier} and {\it RIF} \cite{rif}
Maple packages. When the system has no solution, this fact is detected by
these packages; otherwise the related symmetry of the invariants is
obtained from the output of these packages directly.



\section{Integrable Abel ODE classes found in the literature}
\label{solvable_seeds}

This section is devoted to a compilation of integrable Abel ODE classes
found in the literature. The compilation is not intended to be complete, but
it nevertheless covers various of the usual references; mainly Kamke's and
Murphy's books \cite{kamke,murphy}, and the original works by Abel,
Liouville and others on these subjects \cite{abel,liouville3,appell,halphen}.

One of the noticeable things in these references is that the
presentation of integrable cases lacks a classification in terms of their
invariants. Consequently, many of these ODEs can actually be obtained from
one another by means of \eq{tr}, that is, they belong to the same class.
Since part of this work consisted in writing computer routines addressing
the equivalence problem, we performed this classification, and therefore
present here a more compact collection of integrable Abel ODE {\it
classes}, as opposed to just integrable ODEs. Classes not depending on
parameters are labelled by numbers (e.g., Class 1), while those depending
on parameters are labelled with letters (e.g., Class A). The solutions to
the representatives of these classes are presented altogether in a table in
the Appendix.


While revising the related literature we also noticed that various of the
cases presented in books or papers are in fact particular cases of the
integrable classes presented by Abel, Liouville and Appell in
\cite{abel,liouville3,appell}. In turn the methods they used to obtain new
integrable classes seem to be forgotten or not mentioned elsewhere. So, it
appeared reasonable to start by reviewing and analyzing
selected parts of those works in this section, and then show in the next section how,
starting from these ideas, additional integrable classes can be obtained.

The first large presentation of integrable cases is due to Abel himself in
\cite{abel}. His idea was to consider integrating factors of the form
\begin{equation} \mu={\e^{r(x,y)}} \label{mu_abel} \end{equation}

\ni for ``Abel" equations written in terms of two arbitrary functions $p$
and $q$ as:

\begin{equation}
\Phi \equiv y\y1+p(x)+q'(x)y = 0
\label{abel_0}
\end{equation}

\ni The first non-trivial case discussed in \cite{abel} was found by taking
$r(x,y)$ as quadratic in $y$:
$$\mu={\e^{(\alpha+\beta\,y+{\gamma}\,{y}^{2})}}$$ where $\alpha$, $\beta$
and $\gamma$ are arbitrary functions of $x$. Abel formulated this problem by
applying Euler's operator to the total derivative $\mu \Phi$, obtaining a
system easily solvable for $\alpha$, $\beta$, $\gamma$ and $p$. The
resulting Abel family has non-constant invariant and is shown in Abel's
memoires as depending on one arbitrary function $q(x)$ and two arbitrary
constants $C_i$:

\begin{equation}
y\y1-{\frac {{\it q'}}{2\,C_1\,q+C_2}}+{\it q'}\,y
\label{draft_1000}
\end{equation}


\ni (for the corresponding integrating factor see \cite{abel}). Now, for the
purpose of building computer routines addressing the equivalence problem, it
is crucial to determine whether or not a given class depends on
parameters since, as explained in \se{parameterized_seed_method}, in such a
case the formulation of that problem is much more difficult. In
the case of \eq{draft_1000}, the two parameters $C_i$ and the function
$q(x)$, can be removed by first converting the ODE to first kind using
$y(x)=1/v(x)$, and then employing
a transformation of the form \eq{tr}: $\{x=F(t),v(x)=u(t)\sqrt{-2 C_1}\}$,
with $F$ implicitly defined by $2 C_1 q(F)- t \sqrt{-2 C_1}+C_2=0$,
arriving at a representative of the class simpler
than \eq{draft_1000},

\begin{equation}
\y1={\frac {{y}^{3}}{x}}+{y}^{2}
\label{1000}
\end{equation}

%

\ni and showing that this class does not depend on parameters. It is then
easy to verify that \eq{1000} is a particular case of a parameterized
class\footnote{\eq{1000} is obtained from \eq{class_D} taking $C=0$ and
changing variables $\{x=i\,t,y=i\,u(t)\}$.} derived from Appell's work
\cite{appell}.

The next integrable case shown by Abel is obtained by considering for
\eq{abel_0} an integrating factor of the form
$\mu=\exp{({1}/{(\alpha+\beta\,y)})}$. Proceeding as in the previous case,
Abel arrived at another integrable ODE class with non-constant invariant,
which however (see \cite{liouville3}) is a particular member of the
parameterized class \eq{class_A} shown by Abel in the same paper.

\ni {\underline{\it Constant Invariant case}}

\ni Abel then considered an integrating factor of the form $ \mu=\left
(\alpha+\beta\,y\right )^{n}$. This ansatz does not lead to a non-constant
invariant family. However, this is the first presentation we have found of a
method for the {\it constant} invariant case. Liouville, and others after him,
rediscovered this method, presented in Kamke as due to M. Chini
\cite{chini}, and in Murphy's book as a change of variables mapping an Abel
ODE into a separable one. A recent discussion of the symmetries of
this {\it constant} invariant problem is found in \cite{schwarz}.

%

\ni {\underline{\it Class ``A" depending on one arbitrary parameter}}

\ni The next ansatz considered by Abel was
\begin{equation}
\mu=\left (A+y\right )^{a}\left (B+y\right )^{b}y
\label{mu_ab}
\end{equation}

\ni where $A(x)$ and $B(x)$ are arbitrary functions and $a$ and $b$ are
arbitrary constants. By taking $b=-a$ Abel showed that a tractable
integrable case results:

\begin{equation}
y\y1
+ \fr {{\it q'}} {4\,q}
    \left (\left (q+2\,{\frac {C_1}{q}}\right )^{2}
    -
    {\frac {{q}^{2}}{{a}^{2}}}\right )
    +{\it q'}\,y
= 0
\label{abel_class_A}
\end{equation}

\ni The arbitrary function $q(x)$ can be removed together with the constant
$C_1$ by rewriting this ODE in first kind format, and then appropriately
choosing \FPQ\ in \eq{tr}; so that a simpler representative of this class
depending on only one parameter ``$\alpha$", is given by\footnote{A representative
of the same class of \eq{class_A} is shown in \cite{liouville3} as
$\y1=\fr{4}{9\,{x}^{3}}\, {\left(\left ({x}^{2}+1\right )^{
2}-c{x}^{4}\right ){y}^{3}}+\fr{4\,{y}^{2}}{3}$}


\begin{equation}
\y1=\left (\alpha\,x+\fr{1}{x}+\fr{1}{x^{3}}\right ){y}^{3}+{y}^{2}
\label{class_A}
\end{equation}

\ni {\underline{\it Class 1}}


In \cite{halphen}, Halphen noted a connection between doubly-periodic
elliptic functions and the Abel type ODE

\begin{equation}
\y1={\frac {3\,y\left (1+y\right )-4\,x}{x\left (8\,y-1\right )}}
\label{class_1}
\end{equation}

\ni which transforms into itself under infinitely many rational changes of
variables, from where he was able to determine both a parametric and an
algebraic solution for it (see the Appendix).

\ni {\underline{\it Class 2}}

\ni In a paper by Liouville \cite{liouville3} mostly dedicated to Abel
equations, he discussed the integrable cases known at that time (1903), and
presented some new ones. Liouville reviewed Abel's work and considered for
\eq{abel_0} an integrating factor of the form \eq{mu_abel} with $r(x,y)$
cubic in $y$, arriving at the integrable family
$\y1=6\,a\,x{y}^{2}+3\,a\,{y}^{3}$, depending on a parameter $a$. This
parameter however can be removed by changing variables as in $
\{y=-{{u(t)}/{\sqrt [3]{3\,a}}},x={ {t}/{\sqrt [3]{3 \,a}}}\}$
arriving at the integrable class free of parameters represented by


\begin{equation}
\y1=-2\,{y}^{2}x+{y}^{3}
\label{class_2}
\end{equation}

\ni {\underline{\it Class ``B" depending on one arbitrary parameter}}

\ni As a generalization of \eq{class_2}, in \cite{liouville3} Liouville also
presented the parameterized family

\begin{equation}
\y1+\left (3\,m{x}^{2}+4\,{m}^{2}x+n\right ){y}^{3}+3\,x{y}^{2}=0
\label{seed_B}
\end{equation}

\ni written in terms of two parameters $m$ and $n$ and which can be mapped
into a Riccati ODE solvable in terms of special functions. \eq{class_2} is a
member of the class represented by \eq{seed_B} after setting $m=0$. However,
when $m = 0$, $n$ can be removed from \eq{seed_B} by changing variables
$\left \{x=t\sqrt [3]{n},\ y(x)={{t\,u(t)}/ {{{n}}^{2/3}} }\right \}$,
leading to a class without parameters - actually represented by
\eq{class_2}. In turn, when $m \neq 0$, $m$ and $n$ can be ``merged" by
changing variables $\left \{y={{u(t)}/{{m}^{2}}},x=mt\right \}$ and
introducing a new parameter $a=n/{m}^{3}$, resulting in


\begin{equation}
\y1= -\left (3\,{x}^{2}+4\,x+a\right ){y}^{3}-3\,x{y}^{2}
\label{class_B}
\end{equation}

\ni In summary, \eq{seed_B} is not a full 2-parameter class, but instead
two classes represented by Eqs.(\ref{class_2}) and (\ref{class_B}),
respectively depending on zero and one parameters.
A simpler representative for this class and its solution are found in the Appendix.

\ni {\underline{\it Class 3}}

Still in \cite{liouville3} Liouville pointed out that by interchanging the
role between the dependent and independent variables in \eq{class_2} one
arrives at a different Abel integrable class. After rewriting this resulting
ODE in first kind format and performing a change of variables of the form
\eq{tr}, a simpler representative of this integrable class is given by

\begin{equation}
\y1={\frac {{y}^{3}}{4\,{x}^{2}}}-{y}^{2}
\label{class_3}
\end{equation}

\subsection{\underline{\it Integrable Abel ODE classes shown in Kamke and
some others books}}

One of the most well known collection of (69) Abel ODEs is the one shown in
Kamke's book. This collection however makes no distinction between constant
or non-constant invariant cases, presents ODEs of the same class as
different, and does not discuss what would be the representative for each
class depending on the least number of parameters. A first classification for
these Abel ODEs is then given by\footnote{In this classification, by ``too
general" we mean: these ODEs cannot be solved without restricting the
example to a concrete particular case. We excluded ODEs - like those numbered 230 and 232 - which are already of Bernoulli type. We note also that the ODEs shown in
Kamke without solution can all be transformed into an Emden type second
order ODE shown in Kamke as 6.74, for which only a general discussion is
presented. In turn, a detailed discussion on the integrable cases of Emden
type ODEs is found in \cite{green}.}:

{\begin{center} {\footnotesize
\begin{tabular}{|p{1.8 in}|p{4.3 in}|}
\hline
Classification  &  ODE numbers as in Kamke's book \\
\hline 
4 are too general &
50, 219, 250, 269
 \\
\hline
40 constant invariant &
38, 41, 46, 49, 51, 188, 204, 213, 214, 215, 216, 218, 221, 222,
223, 224, 225, 226, 227, 228, 229, 231, 236, 238, 239, 243, 244, 245, 246,
247, 248, 249, 251, 252, 254, 255, 260, 261, 262, 264
\\
\hline
24 non-constant invariant &
36, 37, 40, 42, 43, 45, 47, 48, 111, 145, 146, 147, 151, 169, 185, 203, 205, 206, 234,
235, 237, 253, 257, 265
 \\
\hline
10 shown without solution &
40, 47, 48, 203, 205, 206, 234, 237, 253, 265
 \\
\hline
\multicolumn{2}{c}{Table 1. First classification for the 69 Abel ODEs shown
in Kamke's book.}
\end{tabular}}
\end{center}}

\ni As mentioned, all constant invariant ODEs can systematically be
transformed into separable ODEs (see for instance Murphy's book), so that
the interesting subset is the one comprising 24 ODEs having non-constant
invariants. We note also that 10 of these 24 ODEs are shown in the book
without a solution, and in fact we were unable to solve any of 203, 205,
206, 234, 253 or 265, so that the number of integrable cases for us is 18.

From these 18 ODEs (and hence from the 69 Abel type Kamke's examples), only
four - those numbered: 47, 185, 235 and 237- would really lead to additional
integrable classes with respect to those presented in the works by Abel,
Liouville and Appell. We note however that the examples 47, 185 and 237 are
all members of Class ``C" (see \eq{class_C}), which can be derived from the
work by Abel \cite{abel} - even when it was not presented in the original
work. So that the number of additional integrable classes presented in Kamke
reduces to one, represented by the example 235. The classification and
details are as follows.

\ni {\underline{\it Class 4}}

\begin{equation}
\left (xy+a\right )\y1+b\,y=0
\end{equation}

\ni This ODE (K 1.235) is presented in Kamke in terms of two arbitrary
parameters $\{a,\,b\}$; then, a change of variables which
transforms it into a linear ODE is shown. A simpler representative of this class -
not depending on parameters - can be obtained by rewriting this equation
in first kind format via $\{x=t,y={\frac {1}{t\,u(t)}}-{\frac {a}{t}}\}$ and
then changing variables $\{x={\frac {a}{t\,b}},y={\frac {t\,u(t)}{a}}\}$,
leading to

\begin{equation}
\y1={y}^{3}-{\frac {\left (x+1\right )}{x}}{y}^{2}
\label{class_4}
\end{equation}

\ni {\underline{\it Comments on Kamke's example 47}}  


For the ODE 
\begin{equation}
\y1-a\left ({x}^{n}-x\right ){y}^{3}-{y}^{2}=0
\label{k47}
\end{equation}

\ni presented in Kamke as K 1.47, there is no solution shown in the book,
but instead a suggestion of transforming the ODE into a second order one.
We followed that suggestion and then ran a symmetry analysis, noticing that
the resulting ODE will have two point symmetries if either $ \{a=-{\frac
{2\,n+2}{9+6\,n+{n}^{2}}}\}$ or $\{n=2,\, a={\frac {6}{25}}\}$, leading to
two integrable classes not shown in the book. In the former case, from
\eq{k47}, we arrive at

\begin{equation}
\y1+{\frac {\left (2\,n+2\right )\left ({x}^{n}-x\right ){y}^{3}}{9+{
n}^{2}+6\,n}}-{y}^{2}=0
\label{class_E}
\end{equation}

\ni However, this ODE can be transformed into \eq{class_C} by changing
variables $\{x={t}^{\fr{2}{1-n}}, y=-u(t) \fr{n+3}{2}{t}^{{\frac
{n+1}{n-1}}} \}$ followed by $n={\frac {a+2}{a-2}}$, so that it belongs to
Class C. In the same line, taking $\{n=2,\, a={\frac {6}{25}}\}$ in
\eq{k47}, and changing variables $\{x={\frac
{{t}^{2}-1}{{t}^{2}}},y=5/2\,u(t){t}^{3}\}$ one arrives at \eq{class_C} with
$a=6$, so that this second branch of \eq{k47} is also a member of Class C.

%

\ni {\underline{\it Comments on Kamke's example 237}}  

\begin{equation}
x\left (y+a\right )\y1+b\,y+c\,x = 0
\label{237}
\end{equation}

\ni This ODE (K 1.237) depending on three arbitrary parameters
$\{a,\,b,\,c\}$, is presented in the book without a solution. We note
however that changing $\{x \rightarrow y,\ y \rightarrow x\}$ leads to an
ODE also of Abel type and in second kind format. Converting the latter to
first kind format via $\{x=t,y={\frac {1}{c\,u(t)}}-{\frac {b\,t}{c}} \}$,
replacing $y \rightarrow y'$ and running a symmetry analysis, the resulting
second order ODE has two symmetries when $a=-2\,b$, leading to an integrable
case. Introducing $a=-2\,b$ into \eq{237}, rewriting it in first kind format
via $\{x=t,y=-{\frac {1}{t\,u(t)}}+2\,b\}$ and changing variables
$\{x=-{\frac {{b}^{2}\left (t+4\right )}{2\,c}},y={\frac {
2\,c\,u(t)}{{b}^{3}\left (t+4\right )}}\}$ leads to a simpler representative
of the class not depending on any parameters:

\begin{equation}
\y1={\frac {-x\,{y}^{3}+2\,{y}^{2}}{2\,(x+4)}}
\label{class_310}
\end{equation}

\ni However, by changing variables $\
\{x=4\,(1-{t}^{2})/{t}^{2},y=-u(t)t/2\}$ one arrives at \eq{class_C} again,
this time with $a=-1/2$, so that \eq{class_310} is also member of Class C.



A classification for all these 18 non-constant invariant Kamke examples is
then as follows\footnote{Equations K.1.47, K.1.48 and K.1.237 belong to
Class C for infinitely many - however particular - values of one of the two
parameters (see \eq{class_E}); we don't know their solution for other values.}

{\begin{center} {\footnotesize
\begin{tabular}{                     
|c|c|c|c|c|c|c|
}
\hline
Class 2 & Class 3 & Class 4 & Class A & Class B & Class C & Class D  \\
\hline
36, 40 & 145, 147 & 235 & 257 & 42, 43 & 45, 47, 48, 151, 185, 237 & 37, 111, 146, 169 \\
\hline
\multicolumn{7}{c}{Table 2. Classification for the 18 non-constant invariant
solvable Abel ODEs in Kamke's book.}

\end{tabular}}
\end{center}}

\ni where classes C and D are defined in \se{new_seeds}. In summary, all
but one of Kamke's 58 solvable examples (18 non-constant invariant + 40
constant invariant) are particular cases of the integrable classes
presented by Abel, Liouville and Appell in \cite{abel,liouville3,appell}, or
can be derived from there (those belonging to Classes C and D).

Another collection of Abel ODEs is found in the book by Murphy
\cite{murphy}. After selecting those examples not having a constant invariant
and for which a solution is shown in the book, we arrived at a set
of nine ODEs, numbered in the book as: 78, 79, 80, 86, 275, 304, 345, 383
and 593. None of these ODEs represent an additional integrable class; their
distribution among the classes discussed in this work is as follows

{\begin{center} {\footnotesize
\begin{tabular}{                     
|c|c|c|c|c|
}
\hline
Class 2 & Class 3 & Class B & Class C & Class D  \\
\hline
78, 80  & 275     &  86    & 304, 383, 593 & 79, 345 \\
\hline
\multicolumn{5}{c}{Table 3. Classification for the non-constant invariant
solvable Abel ODEs in Murphy's book.}

\end{tabular}}
\end{center}}

A wider collection of Abel ODEs than the one shown in Kamke's book is found
in the book by Polyanin and Zaitsev \cite{green}. This book is rather new
(1995) and covers a vast number of integrable ODE problems which we have not
found in other books, hence making the examples attractive. On the other
hand the Abel ODEs shown there are classified not according to their
invariants but according to their form, and the origin of their solutions is
not given. Apart from a main section consisting of four tables (82 Abel ODEs
- all derived from four basic ones), the book contains other sections
illustrating mappings between Abel and higher order ODEs. The quantity of
examples is large and the computational routines we prepared for the
equivalence problem are not yet covering in full the case in which the
parameters of the class may assume {\it symbolic} values. As a result we still
don't have a way to solve the equivalence problem for the whole set of
integrable classes presented in \cite{green}. Our analysis of these Abel ODEs of
\cite{green} is then still incomplete; consequently we restricted the
presentation here to just a sample, constituted by the ODEs of the first of
these four tables. These are 20 ODEs obtained from 

\begin{equation} y\y1-y=sx+A{x}^{m} \end{equation}

\ni by giving particular values to the parameters $m$ and $s$ ($A$ is kept
arbitrary). These ODEs appear in section 1.3.1 of \cite{green} under the
numbers: 1, 2, 10, 16, 19, 22, 23, 26, 27, 30, 32, 33, 45, 46, 47, 48, 53,
54, 55 and 56. We were not able to classify those numbered 27, 20, 48, 55
and 56. The distribution of the remaining ODEs, in the classes
discussed in this work, is as follows:

{\begin{center} {\footnotesize
\begin{tabular}{                     
|c|c|c|c|c|c|
}
\hline
Constant invariant & Class 1 & Class 2 & Class 3 & Class C & Class D   \\
\hline
1, 2, 26
&         
23
&        
32
&        
33
%
%
&        
10, 19, 22, 45, 46, 47, 53, 54
&        
16


\\
\hline
\multicolumn{6}{c}{Table 4. Classification for 15 of the 20 Abel ODE
examples of Table 1.1 of \cite{green}.}

\end{tabular}}
\end{center}}

\section{New integrable Abel ODE classes derived from previous works}
\label{new_seeds}

\ni {\underline{\it Class ``C" depending on one arbitrary parameter}}

The form of the integrating factor studied by Abel actually leads to
other integrable cases not mentioned in the original work \cite{abel}. One
of them is obtained by taking $b=a$ in \eq{mu_ab}, resulting in the ODE
family\footnote{$n$ in \eq{abel_b_eq_a} is related to $a$ in \eq{mu_ab} by
$n=-1/(2a+1)$}

\begin{equation}
y\y1-{\it q'}\,y
-\fr
{{\it q'}\,{n}^{2}\left (-{\frac {q}{n}}+{C_1}^{2}
\left ({\frac {q}{n}}\right )^{2\,n-1}\right )}
{\left (n+1\right )^{2}} = 0
\label{abel_b_eq_a}
\end{equation}

\ni where $n \neq -1$. The function $q(x)$ and the parameter $C_1$ can be
removed as done with \eq{abel_class_A}, leading to

\begin{equation}
\y1=n\left (x-{x}^{2\,n-1}\right ){y}^{3} - \left (n+1\right ){y}^{2}
\end{equation}

\ni which is turned exact by means of the integrating factor

\begin{equation}
\mu = 
\fr
{\left (1+\left (\left ({x}^{2}-{x}^{2\,n}\right )y
-2\,x\right )y\right )^{-{\frac {n+1}{2\,n}}}}
{{y}^{{\frac {2\,n-1}{n}}}}
\label{mu_class_C_draft}
\end{equation}

\ni A simpler representative of this class is obtained by
changing variables $\{y=u(t){t}^{{\frac {n}{n-1}}},x={t}^{\fr{1}{1-n}}\}$,
then introducing a new parameter by means of $n={\frac
{\alpha}{\alpha-2}}$, arriving at

\begin{equation}
\y1={\frac {\alpha\left (1-{x}^{2}\right ){y}^{3}}{2\,x}}+\left (\alpha-1
\right ){y}^{2}-{\frac {\alpha y}{2\,x}}
\label{class_C}
\end{equation}

\ni Taking into account \eq{mu_class_C_draft}, an implicit solution for
this class is given by
\begin{equation}
C_1+\fr{\alpha}{x}\left (1-{\frac {\left (1-xy\right )^{2}}{{y}^{2}}}\right )^{{1/\alpha}}
-2\,{\int}^{^{\frac {1-xy}{y}}}\!\left (1-{z}^{2}\right )^{
{\frac {1-\alpha}{\alpha}}}{dz}=0
\end{equation}

\ni {\underline{\it Class ``D" depending on one arbitrary parameter}}

In \cite{appell}, Appell showed a series of integrable cases derived from
the solutions to
\begin{equation}
u'=A(u)+B(u)\,t
\label{apple_trick}
\end{equation}

\ni By changing variables $\{t=\fr{1}{y}-\fr{A(x)}{B(x)},\,u=x \}$, this ODE is
transformed into the Abel ODE

\begin{equation}
\y1= -{\frac {{y}^{3}}{B}}
- \left ({\frac {A}{B}}\right )' {y}^{2} 
\label{appell_abel}
\end{equation}

\ni where $A$ and $B$ are now functions of $x$. Any particular
$\{A,\,B\}$ leading to a solvable case in \eq{apple_trick} will then also
lead to an integrable Abel ODE \eq{appell_abel}. Among the choices for
$\{A,\,B\}$ considered in \cite{appell} - such that \eq{apple_trick} results
linear, homogeneous, or of Riccati type - only this mapping into Riccati
type leads to something new. This case is obtained by taking

\begin{equation}
A=ax^2+bx+c,\ \ \ \ B=\alpha x^2+\beta x + \gamma
\end{equation}

\ni The related Abel ODE family, depending on six parameters
$\{a,\,b,\,c,\,\alpha,\,\beta,\,\gamma\}$, is given by

\begin{equation}
\y1=-{\frac {{y}^{3}}{\alpha\,{x}^{2}+\beta\,x+{\gamma}}}
-
{y}^{2}\,
\fr{d}{dx}\left ({\frac {a{x}^{2}+bx+c}{\alpha\,{x}^{2}+\beta\,x+{\gamma}}}\right )
\label{class_D_5p}
\end{equation}

\ni and its solution could be expressed in terms of the solution to the
Riccati ODE

\begin{equation}
\y1=\left (a+\alpha x\right ){y}^{2}+\left (b+\beta\,x\right )y+c+{\gamma} x
\label{appell_riccati}
\end{equation}

\ni However, we were not able to solve this Riccati ODE for arbitrary values
of the six parameters involved and in \cite{appell} there is no indication
of how that could be done. The alternative we then investigated is to
consider the second order ODE obtained by replacing $y=y'$ in
\eq{appell_riccati}. That ODE has {\it two} point symmetries if and
only if $\alpha = 0$. With these symmetries we were able to solve that
second order ODE, and hence \eq{appell_riccati} when $\alpha = 0$.
Concerning the related Abel family \eq{class_D_5p} - now depending on five
parameters - an appropriate change of variables of the form \eq{tr}

\begin{equation}
\left \{x={\frac {t\sqrt {\beta}}{a}}-{\frac {{\gamma}}{\beta}},\ \ y=
\sqrt {\beta}\,u(t)\right \}
\end{equation}

\ni followed by the introduction of a new parameter $C$ by means of


\begin{equation}
C=-{\frac {\left ({\beta}^{2}\,c+\alpha\,{{\gamma}}^{2}\right )a
-\alpha\,\beta\,{\gamma}\,b}{{\beta}^{2}}}
\end{equation}

\ni transforms \eq{class_D_5p} into a simpler representative for the class


\begin{equation}
\y1=-{\frac {{y}^{3}}{x}}-{\frac {\left (C+{x}^{2}\right ){y}^{2}}{{x}
^{2}}}
\label{class_D}
\end{equation}

\ni also showing that this class depends not on five but on one
parameter\footnote{We note that in this process we have made two implicit
assumptions: $a\neq 0$ and $\beta\neq 0$. To assure that the cases in
\eq{class_D_5p} are covered by \eq{class_D} we then also considered $a=0$
and $\beta=0$ separately, arriving at ODEs respectively members of the
classes represented by \eq{class_D} and \eq{class_2}.}. It appeared of value
to us also to determine the number of parameters on which
\eq{class_D_5p} depends in the general case, that is {\it before} taking
$\alpha=0$. For that purpose we searched for the appropriate changes of
variables of the form \eq{tr} which would remove as many as possible of these
parameters, requiring that both the change and its inverse are finite. We then
considered the branches which become infinite for some particular values of
the parameters $\{a,\,b,\,c,\,\alpha,\,\beta,\,\gamma\}$ entering the
transformations found. The results are summarized as follows. If all these
parameters are different from zero, introducing new parameters $\{A,\, B,\,
C,\, G\}$ by means of

\begin{eqnarray}
\alpha & = & {\frac {{\beta}^{2}+4\,{A}^{4}}{4\,\gamma}}
\nonumber
\\*[.13 in]
b & = & {\frac {8\,\beta\,{\gamma}^{2}\,a\,{A}^{2}B+C}{2\,\gamma\,{A}^{2}B\left ({\beta}^{2}+4
\,{A}^{4}\right )}}
\nonumber
\\*[.13 in]
c & = &{\frac {{A}^{2}B\,C+16\,{\gamma}^{2}\,a\,{A}^{6}B+
\beta\,C+4\,{\beta}^{2}{\gamma}^{2}\,a\,{A}^{2}B}
{{A}^{2}B\left ( {\beta}^{2}+4\,{A}^{4} \right )^2}}
\nonumber
\\*[.13 in]
\gamma & = &{\frac {C}{2\,{A}^{3}B\,G\left ({\beta}^{2}+4\,{A}^{4}\right )}}
\label{new_params}
\end{eqnarray}

\ni followed by changing variables $\{x={\frac {C\left
(2\,t{A}^{2}-\beta\right )}{\left ({ \beta}^{2}+4\,{A}^{4}\right
)^{2}{A}^{3}B\,G}},\, y=u(t)A\}$ in the six-parameter \eq{class_D_5p}, one
arrives at a 2-parameter representative for the same class

\begin{equation}
\y1=-{\frac {{y}^{3}}{{x}^{2}+1}}+{\frac {G\left (B\,x+{x}^{2}-1\right 
){y}^{2}}{\left ({x}^{2}+1\right )^{2}}}
\label{class_D_2_params}
\end{equation}

\ni Now the case $\alpha=0$ was already shown to lead to \eq{class_D}, and
all the other possible branches obtained from \eq{class_D_5p} by taking some of the other parameters equal to zero lead
either to constant invariant families, or to members of the classes already
discussed in this work\footnote{There is a special case, when $b=4\,{\frac
{{ \gamma}\,\beta\,a}{{\beta}^{2}+4\,{A}^{4}}}$, where the resulting Abel
ODE can only be obtained from \eq{class_D_2_params} by taking appropriate
limits.}

\ni {\underline{\it Three new classes not depending on parameters}}

While analyzing the works \cite{abel,liouville3,appell} and Kamke's
examples, a large number of symbolic experiments were performed, sometimes
leading to intermediate results which with a bit more work appeared to be
new integrable classes by themselves. This happened three times,
resulting in classes 5, 6 and 7, for which representatives and solutions
are given as follows:

\ni {\underline{\it Class 5}}

\begin{equation}
\y1=-{\frac {\left (2\,x+3\right )\left (x+1\right ){y}^{3}}
{2\,{x}^{5}}}+{\frac {\left (5\,x+8\right ){y}^{2}}{2\,{x}^{3}}}
\label{class_5}
\end{equation}

\ni Solution:
\begin{equation}
C_1+{\frac {\sqrt {A}}{\sqrt [4]{4\,{\frac {\left (x+1\right )^{2}}
{{x}^{2}\,A}}+1}}}+{\int}^{2\,{\frac {x+1}{x\sqrt {A}}}}\!\left ({z
}^{2}+1\right )^{-5/4}{dz}=0
\end{equation}

\ni where $A=\fr{4}{y}-\fr{10}{x}-\fr{6}{x^{2}}-4$.

\ni {\underline{\it Class 6}}

\begin{equation}
\y1=-{\frac {{y}^{3}}{{x}^{2}\left (x-1\right )^{2}}}+{\frac {\left (
1-x-{x}^{2}
\right ){y}^{2}}{{x}^{2}\left (x-1\right )^{2}}}
\label{class_6}
\end{equation}

\ni Solution:
\begin{equation}
C_1-{\rm Ei}\l(1,{\frac {y+{x}^{2}-x}{xy\left (x-1\right )}}\r)
+
\fr
{\left (x-1\right )\,y\,{\e^{{\frac {x-y-{x}^{2}}{xy\left (x-1\right )}}}}}
{x-1+y}
=0
\end{equation}

\ni where ${\rm Ei}(n,x)=\displaystyle {\int}_{\!\!1}^{^{\infty }}{\frac
{{\e^{-xt}}}{{t}^{n}}}{dt}$ is the exponential integral.

\ni {\underline{\it Class 7}}

\begin{equation}
%
\y1={\frac {\left (4\,{x}^{4}+5\,{x}^{2}+1\right ){y}^{3}}{2\,{x}^{3
}}}+{y}^{2}
+{\frac {\left (1- 4\,{x}^{2}\right )y}{2\,x\left ({x}^{2}
+1\right )}}
\label{class_7}
\end{equation}

\ni Solution:
\begin{equation}
%
C_1
+ 2\,{\frac {x+A}{\sqrt [4]{{A}^{2}+1}\left (Ax-1\right )}}
+\int ^{A}\!\left ({z}^{2}+1\right )^{-5/4}{dz}=0
\end{equation}

\ni where $
\displaystyle
A={\frac {x - 2\,y{x}^{4}-3\,y{x}^{2}-y}{x\left (x+y{x}^{2}+y\right )}}
$

\section{Computer algebra routines, tests and performance}
\label{performance}

The two itemized algorithms described in sections \ref{equivalence} and
\ref{parameterized_seed_method} for solving the equivalence problem
between two given Abel ODEs were implemented in Maple R5, in the framework
of the ODEtools package \cite{odetools}. The implementation consists of
various routines, mainly accomplishing the following:

\begin{enumerate}

\item determine whether a given Abel ODE belongs to one of the solvable
classes described in the previous sections; in doing that, determine also
the function $F(t)$ entering \eq{tr} and the value of the parameters in the
case of a parameterized class;

\item use that information to determine the functions $P(t)$ and $Q(t)$
entering \eq{tr} and return a solution to the given ODE by means of
changing variables in the solution available for the representative of the
class (see the Appendix).

\end{enumerate}

\subsection{Representatives with simpler invariants for Classes A, C and D}
\label{simpler_invariants}

While preparing the computational routines being presented, we noticed
that: if on the one hand the solutions to the representatives for the
parameterized classes A, C and D (Eqs.(\ref{class_A}, \ref{class_C},
\ref{class_D})) can be expressed in a relatively simple manner (see the
Appendix), on the other hand, for each of these representatives, the form
of the three invariants entering \eq{system} is much simpler if an
appropriate redefinition of the class parameter followed by a change of
variables of the form \eq{tr} is performed. In turn, the complexity of
these invariants is a relevant issue for a computer algebra implementation
since simpler invariants lead to simpler GCD and resultant computations.

In the case of Class A (\eq{class_A}),  redefining the parameter as $\alpha
= 5/36-\kappa/3$ - where $\kappa$ is the new parameter - and changing
variables $ \{x={\frac {\sqrt {12\,t-6}}{2\,t-1}},y={\frac {\left (2\,t-1
\right )u(t)}{\sqrt {12\,t-6}}} \}$ lead to the class representative

\begin{equation}
\y1=\fr{\l({\left (3\,\kappa-2\,x-{x}^{2}\right ){y}^{3}}-9\,y^2-9\,y \r)}{9\,(2\,x-1)}
\label{new_class_A}
\end{equation}

\ni and the first invariant $s_5^3/s_3^5$ for it is

\begin{equation}
{\frac {\left (9\,{\kappa}^{2}+\left (30\,{x}^{2}-6\,x+3\right )\kappa-6\,
{x}^{3}+5\,{x}^{4}+7\,{x}^{2}\right )^{3}}{9\,\left (\kappa+{x}^{2}\right )^{5
}}}
\label{new_invariant_A_1}
\end{equation}

\ni as opposed to

\begin{equation}
729\,{\frac {\left (27\,{\alpha}^{2}{x}^{8}
-\left (72\,{x}^{6}+270\,{x}^{4}+15\,{x}^{8}\right )
\alpha+54\,{x}^{4}+36\,{x}^{2}+135+15\,{x}^
{6}+2\,{x}^{8}\right )^{3}}{{x}^{4}\left (2\,{x}^{4}-9\,\alpha\,{x}^{
4}+9\,{x}^{2}+27\right )^{5}}}
\label{old_invariant_A_1}
\end{equation}

\ni which is the same first invariant but calculated for \eq{class_A}.
Actually a measure of the Maple computational {\it length} of the three
invariants $[s_5^3/s_3^5,\, s_3 s_7/s_5^2,\, s_5 s_7/s_3^4]$ entering
\eq{system} shows the values [99, 215, 223] when calculated on
\eq{new_class_A} above and [146, 354, 356] when calculated on
\eq{class_A}. Also, the degrees in $x$ of both numerator and denominator
of \eq{new_invariant_A_1} are lower than the corresponding degrees of
\eq{old_invariant_A_1}.

A similar situation happens with Class C (\eq{class_C}), where redefining
the parameter as $\alpha={\frac {\kappa-3}{\kappa}}$ and changing
variables $\left \{x={\frac {\sqrt {\left (\kappa+1\right ) \left (3-\kappa\right
)\left ( 3\,t+4\right )\,t}}{\left (\kappa+1\right )t}},y={\frac
{{t}^{2}u(t)\kappa\left ( \left (3\,t+4\right )\kappa+3\,t+4\right )}{\sqrt { \left
(\kappa+1\right )\left (3-\kappa \right )\left (3\,t+4\right )\,t}}}\right \} $ lead
to a class representative a bit more complicated than \eq{class_C}:

\begin{equation}
\y1=
4\,\left ((\kappa-2)\,x+\kappa-3\right )\kappa\,x\,{y}^{3}+6\,{y
}^{2}-{\frac {\left (6\kappa\,x+5\,\kappa+3\right )y}{\left (3\,x+4
\right )\kappa\,x}}
\label{new_class_C}
\end{equation}

\ni for which, however, the computational {\it length} of the three
invariants $[s_5^3/s_3^5,\, s_3 s_7/s_5^2,\, s_5 s_7/s_3^4]$ is [110, 273,
305] as opposed to [173, 382, 412] when calculated for \eq{class_C}. The
degrees with respect to $x$ of the numerators and denominators of the
invariants of \eq{new_class_C} are also lower than those of the invariants
of \eq{class_C}.

\ni The same reduction in the complexity of the invariants is achieved for
class D (\eq{class_D}) by redefining the parameter via $C={\frac
{9}{8}}\,\sqrt {\kappa}$ and changing variables $\{x={\frac { 3\,\sqrt
{2\,\left (1-{t}^{2}\kappa \right )\,\sqrt {\kappa}}}{4\,(1-t\,\sqrt
{\kappa})}},\, y=\sqrt {2\,\left (1-{t}^{2}\kappa\right )\,\sqrt
{\kappa}}\ u(t)\}$, leading to

\begin{equation}
\y1={\frac {\left (2\,{x}^{2}\kappa-2\right )\kappa\,{y}^{3}-3\,{y}^{
2}\kappa+\kappa\,xy}{1-\kappa\,{x}^{2}}}
\label{new_class_D}
\end{equation}

\ni for which the first invariant, $s_5^3/s_3^5$, is given
by\footnote{The case $\kappa=0$ is treated separately.}

\begin{equation}
{\frac {\left (\left (2\,{x}^{4}+15\,{x}^{3}\right ){\kappa}^{2}
-\left (4\,{x}^{2}+15\,x-9\right )\kappa+2\right )^{3}}{4\,{\kappa}^{3}
\left (\kappa\,{x}^{3}-x+1\right )^{5}}}
\label{new_invariant_D}
\end{equation}

\ni as opposed to

\begin{equation}
729\,{\frac {\left (C+{x}^{2}\right )^{3}\left (2\,{C}^{4}+8\,{x}^{2}{
C}^{3}+\left (12\,{x}^{4}+15\,{x}^{2}\right ){C}^{2}+8\,{x}^{6}C+2\,{x
}^{8}-15\,{x}^{6}+9\,{x}^{4}\right )^{3}}{\left (2\,{C}^{3}+6\,{C}^{2}
{x}^{2}+\left (9\,{x}^{2}+6\,{x}^{4}\right )C+2\,{x}^{6}-9\,{x}^{4}
\right )^{5}}}
\label{old_invariant_D}
\end{equation}

\ni which is the same first invariant but calculated for the class
representative \eq{class_D}. In \eq{new_invariant_D}, not
only the degrees with respect to $x$ but also those with respect to the
class parameter $\kappa$ are lower than the corresponding degrees in
\eq{old_invariant_D}.

Moreover, for class D this change in the representative of the class also
fixes a problem: the representative \eq{class_D} is itself invariant under
$C \rightarrow -C$ followed by the change of variables

\begin{equation}
\left \{x={\frac {i\,C}{t}},y=i\,u(t)\right \}
\label{class_D_symmetry}
\end{equation}

\ni where $i$ is the imaginary unit, and hence \eq{old_invariant_D} is
invariant under $\{ C \rightarrow -C,\ x \rightarrow i\,C/x \}$.
Consequently, if $\{ C,\, F(t),\, P(t),\,Q(t) \}$ is a solution to the
equivalence problem between a given ODE and \eq{class_D}, then $\{-C,\,
\fr{ {i\,C}}{F(t)},\, i\,P(t),\,Q(t)\}$ is also a solution; this fact
invalidates the use of the interpolation method described in
\se{symbolic_C} with \eq{class_D} since that method can only be used when
the interpolated solution $C(\alpha)$ is unique.

\subsection{Installation}

The programs being presented have been written as one more step in the
development of the {\it ODEtools} Maple package \cite{odetools} and hence
are integrated to it and not distributed separately. To install the new
Abel routines then what is necessary is to install {\it ODEtools} to run
in the Maple environment by putting the related libraries (two files - {\bf
maple.ind} and {\bf maple.lib}) in any directory - say {\it
DE\_libraries\_directory} - then opening Maple, and adding that directory
to Maple's {\it libname} variable via

\begin{verbatim}
> libname := DE_libraries_directory, libname:
\end{verbatim}

\ni where `\verb->-` is the Maple prompt. This instruction automatically
updates Maple's {\bf dsolve} subroutines to make use of the new routines
for solving Abel ODEs - no further steps are required. In this way, the
new Abel ODE routines are automatically used by Maple's {\bf dsolve} when
the input ODE is of Abel type, as well as to solve higher order ODEs when
they can be reduced to first order Abel ODEs members of the classes
discussed in \se{solvable_seeds} and \ref{new_seeds}.

Apart from this integration with {\bf dsolve}, it is also possible to try
{\it just} the routines being presented by giving to {\bf dsolve} the
extra argument \verb-[Abel]-. For example (Kamke's ODE 37)

\begin{verbatim}
> ode[37] := diff(y(x),x)-y(x)^3-a*exp(x)*y(x)^2 = 0;
\end{verbatim}
$$
ode_{37} := \y1-{y}^{3}-a{\e^{x}}{y}^{2}=0
$$

\begin{verbatim}
> dsolve(ode[37], [Abel]);
\end{verbatim}
\begin{equation}
C_1+{\frac {{\e^{-1/2\,\left (a{\e^{x}}+{y}^{-1}\right )^{2}}}}{a{\e^{x}}
}}+\fr{\sqrt {2\pi }}{2}\,
{\rm erf}\l(\fr{\sqrt {2}}{2}\,\left (a{\e^{x}}+{y}^{-1}\right )\r)
=0
\end{equation}

\ni These implicit answers can also be tested in the Maple worksheet,
interactively, using the standard Maple {\bf odetest} command.

Due to the intrinsic complexity of Abel equations, the solution for most
of the solvable classes is expressed in implicit form and in terms of
elliptic integrals and special or hypergeometric functions. Then, to save
the time Maple spends in trying to ``integrate" these integrals or to
``invert" these algebraic expressions it is frequently convenient to call
{\bf dsolve} with the optional extra arguments \verb-'useInt, implicit'-,
meaning: use Maple's inert \verb-Int- and return the solution directly in
implicit form\footnote{The option \verb-implicit- is standard in Maple's
{\bf dsolve} and the implementation of the option \verb-useInt- comes with
the routines for Abel ODEs presented here.}. For example, for Kamke's
185

\begin{verbatim}
> ode[185] := x^7*diff(y(x),x)+2*(x^2+1)*y(x)^3+5*x^3*y(x)^2 = 0
\end{verbatim}
$$
ode_{185} := {x}^{7}\y1+2\,\left ({x}^{2}+1\right ){y}^{3}+5\,{x}^{3}{y}^{2}
=0
$$

\ni the solution obtained using these extra arguments is:
\begin{verbatim}
> dsolve(ode[185], [Abel], useInt, implicit);
\end{verbatim}
\begin{equation}
C_1+{\frac {x}{\sqrt [4]{{A}^{2}+1}}}+1/2\,\int^{A
}\!\left ({z}^{2}+1\right )^{-5/4}{dz} = 0
\end{equation}

\ni where $A={x}^{-1}+{\frac {{x}^{2}}{y}}$. The explicit computation of
the integral above leads to a complicated expression with hypergeometric
functions somehow obscuring the structure of the solution.

Concerning Maple's difficulty in solving the problem when the class
parameter $\C$ involves radicals (see end of \se{numerical_C}), we implemented an environment variable
controlling how {\it hard} the routines will work, in order to avoid
exhausting system resources unless a hard trial is specifically requested.
This environment variable is \verb-_Env_odsolve_Abel_try_hard-, it can be
assigned a positive integer from 1 to 5, and by default it is assigned to
4; the meaning of the possible values is  as follows:

\begin{itemize}

\item if \verb-_Env_odsolve_Abel_try_hard = 1- then the algorithm for
parameterized Abel classes is disabled, and for non-parameterized classes
only a restricted equivalence using $\{x=t, y=P(t)\,u+Q\}$, that is: with
$F(t)=t$ in \eq{tr}, is tried;

\item if \verb-_Env_odsolve_Abel_try_hard = 2- then for non-parameterized
classes a full equivalence using \eq{tr} is tried;

\item if \verb-_Env_odsolve_Abel_try_hard = 3- then the algorithm for
parameterized Abel classes is enabled but only for {\it numerical}
solutions for the class parameter $\C$;

\item if \verb-_Env_odsolve_Abel_try_hard = 4- then for parameterized Abel
classes both a {\it numerical} or a rational interpolation solution for
the class parameter $\C$ are tried (symbolic variables in the coefficients
are allowed but solutions with radicals are not computed) ;

\item if \verb-_Env_odsolve_Abel_try_hard = 5- then the algorithm for
parameterized Abel classes will also compute solutions involving radicals
for the class parameter $\C$.

\end{itemize}

Finally, these routines for Abel ODEs were programmed to provide extensive
run-time information on the computations being performed through the Maple
standard {\bf userinfo \& infolevel} scheme. For example, Kamke's ODE 43
belongs to class B (see Table 2.) and both the way how this is determined
and the explicit values found for $F$, $P$, $Q$, and the parameter entering
\eq{class_B} can be seen by setting the {\bf infolevel} as
follows\footnote{We kept few lines to illustrate the {\it userinfo}
feature, and represented the missing ones by \verb-`....`-.}:

\begin{verbatim}
> infolevel[dsolve] := 4;
> ode[43] := 
\end{verbatim}
$$
ode_{43} := \y1+\left (3\,a{x}^{2}+4\,{a}^{2}x+b\right ){y}^{3}+3\,{y}^{2}x
=0
$$
\begin{verbatim}
> dsolve(ode[43], [Abel], implicit);
                           ....
The relative invariant s3 is: -3*a*x^2+b-2*x^3
The first absolute invariant s5^3/s3^5 is: 108*(12*a^2*x^3-8*a*x*b + ...
The second absolute invariant s3*s7/s5^2 is: 1/3*(3*a*x^2-b+2*x^3) * ...
The third absolute invariant s5*s7/s3^4 is: 9*(372*a^3*x^4+450*a^2*x^5 + ...
                           ....
                           ....
->         ======================================
->             ...checking Abel class B (by Liouville)
Trying a = 0
Trying a = 1
Trying b = 0
Trying b = 1
-> Step 1: checking for a disqualifying factor on F after evaluating x
Trying x =   0
*** No disqualifying factor on F was found ***
-> Step 2: calculating resultants to eliminate F and get candidates for C
*** Candidates for C are   [1, 4, 1/4], ***
-> Step 3: looking for a solution F depending on x
_____________________________
C = 1/4  leads to the solutions  [{F = -1-3/2*x}]
Interpolated candidate for the class parameter C is: C = 1/4
General testing of the candidate C = 1/4 with arbitrary b
Interpolation is still incomplete; trying next value of b
_____________________________
Trying b = 2
                           ....
                           ....
General testing of the candidate C = -3/4*b+1 with arbitrary b
_____________________________
C = -3/4*b+1  leads to the solutions  [{F = -1-3/2*x}]
General test of C = -3/4*b+1 passed OK;
interpolation for b in this level is complete.
                           ....
                           ....
General testing of the candidate C = (-3/4*b+a^3)/a^3 with arbitrary a
_____________________________
C = (-3/4*b+a^3)/a^3  leads to the solutions  [{F = -1/2*(3*x+2*a)/a}]
General test of C = (-3/4*b+a^3)/a^3 passed OK;
interpolation for b in this level is complete.
_____________________________
Value of the Class parameter solving the problem is:
                  C = 1/4*(-3*b+4*a^3)/a^3
Inverse of the transformation solving the problem is:
        {t = -1/2*(3*x+2*a)/a, u(t) = -2/3*a^2*y(x)}

Solution:
\end{verbatim}
\begin{equation}
C_1+
\fr
{\left(\left ({\frac {2\,a+3\,x}{2\,a}}-A_1\right ){\rm K}(A_1 
,-\sqrt {A_2})-\sqrt {A_2}\,{\rm K}(A_1 +1,-\sqrt {A_2})\right)}
{\left(\left ({\frac {2\,a+3\,x}{2\,a}}-A_1\right ){\rm I}(A_1 ,-\sqrt {A_2}
)+\sqrt {A_2}\,{\rm I}(A_1+1,-\sqrt {A_2})\right )}=0
\end{equation}

\ni where $A_{{1}}=\fr{1}{2}\sqrt {{\frac {- 3\,b+4\,{a}^{3}}{{a}^{3}}}}$,
$A_2={\frac {3\,b-4\,{a}^{3}}{4\,{a}^{3}}} +{\frac {\left (2\,a+3\,x\right
)^{2}}{4\,{a}^{2}}}-{\frac {3}{2\,{a}^{2}y}}$, and I and K are
respectively the modified Bessel functions of first and second kind.

\subsection{Tests and Performance}
\label{tests}

The idea was to test these computational routines to confirm the
correctness of the returned solutions as well as to obtain the
classification presented in the previous sections for solvable Abel ODEs.
The first testing arena was the 69 Abel examples found in Kamke plus the 9
solvable examples with non-constant invariant from Murphy's book, plus the
first 20 Abel ODE examples from \cite{green} mentioned at the end of
\se{solvable_seeds} - totaling 98 Abel ODE examples. The routines passed
these tests - the solutions obtained were confirmed to be correct using other
symbolic computation tools interactively - and the resulting
classification is that shown in Tables 1, 2, 3 and 4 of
\se{solvable_seeds}.

The aforementioned test however involves only 40 ``non-constant invariant and
solvable" Abel ODE examples, and does not fully test the new routines. We
have then set up a more thorough test, which can be taken as a test-suite
for other computer algebra implementations of methods for
solving Abel ODEs. The ideas behind this additional test-suite are
summarized as follows:

\indent $\bullet$ Take the representatives of the seven Abel class not
depending on parameters, discussed in \se{solvable_seeds} and
\ref{new_seeds}, and generate with each four more Abel ODEs of the same
class by applying the different types of transformations:

\begin{enumerate}
\label{test_transformations}
\item The general case with $F,\, P$ and $Q$ arbitrary:
\begin{equation}
tr_1 := \{ x = F(t),\ \  y(x) = P(t) u(t) + Q(t)\}
\label{tr_1}
\end{equation}

\item Rational transformation with symbolic coefficients $a$ and $b$:
\begin{equation}
tr_2 := \{ x = \fr{a}{t} + {b\, t},\ \  y(x) = \fr{u(t)}{t} + 1 \}
\label{tr_2}
\end{equation}

\item Non-rational transformation involving an elementary function and symbols:
\begin{equation}
tr_3 := \{ x = \e^t+1 + \fr{t}{a},\ \  y(x) = u(t) + 1 \}
\label{tr_3}
\end{equation}

\item Non-rational transformation involving abstract powers and symbols:
\begin{equation}
tr_4 := \{ x = \fr{a}{t^n} + \fr{t}{b},\ \  y(x) = u(t) + 1 \}
\label{tr_4}
\end{equation}

\end{enumerate}

For example, the first of these four Abel ODEs generated from
\eq{class_1} is given by:

\begin{eqnarray}
{\it y'} & = & {\frac {{\it F'}\,{P}^{2}\left (-3+F\right ){y}^{3}}{8\,F}}
+{\frac {\left (\left (3\,F-9\right )Q-10\right )P{\it F'}\,{y}^
{2}}{8\,F}}
\\*[.3in]
& & 
+\left ({\frac {\left (\left (3\,F-9\right ){Q}^{2}-20\,
Q-3\right ){\it F'}}{8\,F}}-{\frac {{\it P'}}{P}}\right )y+{\frac {
\left (\left (F-3\right ){Q}^{3}-10\,{Q}^{2}-3\,Q\right ){\it F'}}{8\,FP
}}-{\frac {{\it Q'}}{P}}
\nonumber 
\end{eqnarray}

\ni and the other three are obtained from this one by replacing $F$,
$P$ and $Q$ by the values implied by Eqs. (\ref{tr_2}), (\ref{tr_3}) and (\ref{tr_4}).
The scheme just outlined generates 28 more solvable Abel ODE examples with
non-constant invariant (4 per class), and suffices for testing the solving
of classes without parameters.

Concerning classes with parameters:

\indent $\bullet$ Take the representatives of the four Abel classes
depending on parameters discussed in \se{solvable_seeds} and
\ref{simpler_invariants}, generate with each one four more Abel ODEs of
the same class by applying the transformations \eq{tr_2} and \eq{tr_4},
preceded by replacing the single parameter $\C$ entering each class
representative by $\C = 2$ and $\C ={\alpha}/{\beta}$.

This increases by 4 x 2 x 2 = 16 more ODE examples, and the solving of
these examples tests both the scheme for numeric values of the parameter
$\C$ as well as the case where $\C$ is a rational function of other
symbols entering the given ODE.

The time spent by the routines being presented in solving all these 28 +
16 = 44 additional Abel ODEs with non-constant invariant, is summarized in
Tables 5, 6 and 7 as follows:

{\begin{center} {\footnotesize
\begin{tabular}
{|c|c|c|c|c|}
\hline
Class & transf. (i) & transf. (ii) & transf. (iii)  & transf. (iv) \\
\hline 
1;  \eq{class_1}  &  0.255 sec.   &  0.879 sec. &  0.687 sec.  &  1.256  sec.\\
\hline 
2;  \eq{class_2}  &  1.530 sec.   &  1.900 sec. &  1.998 sec.  &  2.620 sec.\\
\hline 
3;  \eq{class_3}   &  0.277 sec.  &  6.760 sec. &  5.623 sec.  &  11.635 sec.\\
\hline 
4;  \eq{class_4}  &   0.456 sec.  &  7.126 sec. &  3.941 sec.  &  15.511 sec.\\
\hline 
5;  \eq{class_5}   &  0.610 sec.  &  2.505 sec. &  3.082 sec.  &  7.877 sec.\\
\hline 
6;  \eq{class_6}   &  1.379 sec. & 120.952 sec. & 45.565 sec.  &  282.103 sec.\\
\hline 
7;  \eq{class_7}   &  0.896 sec. &  15.441 sec. & 24.586 sec.  &  183.557 sec.\\
\hline 
\multicolumn{5}{c}
{Table 5. \it Timings for 28 Abel ODEs with non-constant invariant - 7 classes free of parameters.}\\
\end{tabular} }
\end{center}}

{\begin{center} {\footnotesize
\begin{tabular}
{|c|c|c|}
\hline
\multicolumn{3}{|c|}
{Class parameter $\C=2$} \\
\hline
Class &  transf. (ii) & transf. (iv) \\
\hline 
A;  \eq{new_class_A}  &  25.546 sec. &  41.304 sec.\\
\hline 
B;  \eq{class_B}  &  29.396 sec. & 54.171 sec.\\
\hline 
C;  \eq{new_class_C}  & 18.648  sec. & 41.425  sec.\\
\hline 
D;  \eq{new_class_D}  & 36.641 sec. & 75.962  sec.\\
\hline 
\multicolumn{3}{c}
{Table 6. \it Timings for 8 Abel ODEs with non-constant invariant - ``numeric" class parameter.}\\
\end{tabular} }
\end{center}}

{\begin{center} {\footnotesize
\begin{tabular}
{|c|c|c|}
\hline
\multicolumn{3}{|c|}
{Class parameter $\C=\fr{\alpha}{\beta}$} \\
\hline
Class &  transf. (ii) & transf. (iv) \\
\hline 
A;  \eq{new_class_A}  &  67.076 sec. &  156.425 sec.\\
\hline 
B;  \eq{class_B}  & 174.751  sec. &  491.210 sec.\\
\hline 
C;  \eq{new_class_C}  & 73.940  sec. &  174.761 sec.\\
\hline 
D;  \eq{new_class_D}  &  88.365 sec. & 177.314 sec.\\
\hline 
\multicolumn{3}{c}
{Table 7. \it Timings for 8 Abel ODEs with non-constant invariant - ``symbolic" class parameter.}\\
\end{tabular} }
\end{center}}

Concerning a comparison of performances between the new routines and those
available in other computer algebra systems (CAS), this appeared to us not
justified in this case: roughly speaking none of these CAS return
solutions for Abel ODEs with non-constant invariant. More precisely, from
Mathematica 3.0, Macsyma 2.7, Maple 5.1 and Mupad 4.0, all of them
failed\footnote{For Kamke's ODE 257, Macsyma (2.7) returns a wrong answer
in terms of $y'$.} in solving any of the 18 Kamke's examples shown in Table
2. (and hence in solving any of the 44 ODEs of Tables 5, 6 and 7), except for
Kamke's example 235 - it is an inverse linear ODE - and anyway none of
them solved it after transforming it to first kind format.

\subsection{Performance with the 1\st order ODE examples from Kamke}
\label{first_order}

Although the main purpose of this paper is to present a
computational scheme for finding solutions to Abel ODEs, it is interesting
to see how {\bf odsolve} - the ODE-solver of the ODEtools Maple package
\cite{odetools} - performs with the addition of these new routines. The
performance with all of Kamke's 553 solvable examples\footnote{We
classified as {\it unsolvable} in general Kamke's examples 50, 55, 56, 74,
79, 82, 202, 219, 250, 269, 331, 370, 461, 503 and 576.} after
incorporating the computational routines presented in this paper is: 97\%
are solved. This performance is summarized as follows

{\begin{center} {\footnotesize
\begin{tabular}
{|c|c|c|c|c|}
\hline
 & & & \multicolumn{2}{|c|}{Average time} \\
\cline{4-5}
Degree in \y1 & ODEs & Solved & {\it solved} & {\it fail}  \\
\hline 
1  & 350 & 337 &  3.2 sec. &  12.9 sec. \\
\hline 
2  & 145 & 140 &  8.8 sec. &  61.1 sec. \\
\hline 
3  & 27 & 26 &  7.2 sec. & 17 sec. \\
\hline 
higher  & 31 & 30 &  13.4 sec. & 25.2 sec. \\
\hline 
\hline 
Total:  & 553 & 533 & $\approx$ 6 sec. & $\approx$ 20 sec. \\
\hline 
\multicolumn{5}{c}
{Table 8. \it Kamke's first order ODEs, solved by {\bf odsolve}: $97 \%$}\\
\end{tabular} }
\end{center}}

The number and classification of Kamke's 1\st order ODEs still not solved by
{\bf odsolve} is now:

{\begin{center} {\footnotesize
\begin{tabular}{|l|p{4.1in}|}
\hline
Class      &  Kamke's numbering \\
\hline 
rational    &
452
   \\
\hline
Riccati     &
25
   \\
\hline
NONE        &
80, 81, 83, 87, 121, 128, 340, 367, 395, 460, 506, 510, 543, 572
   \\
\hline 
\multicolumn{2}{c}{Table 9. \it
Kamke's 1\st order solvable ODEs for which {\bf odsolve} fails: $3 \%$}\\
\end{tabular}}
\end{center}}

\ni where the Abel ODEs numbered in Kamke's book as 47, 48, 205, 206, 237 and
265 not presented in the tables above are known to be solvable only for
specific values of their parameters - and for these values {\bf odsolve}
succeeds - and the ODEs 234 and 253 were not included since their solutions
are not shown in the book or known to us.

\section{Conclusions}
\label{conclusions}

In this paper, a first classification, according to invariant theory, of
solvable non-constant invariant Abel ODEs found in the literature, was
presented. Also, a set of new solvable classes, depending on one or no
parameters, derived from the analysis of the works by Abel, Liouville and
Appell \cite{abel,liouville3,appell}, was shown. Computer algebra routines
were then developed, in the framework of the Maple ODEtools package, to
solve - in principle - any member of these classes by solving its related
equivalence problem. The result is a concrete new tool for solving Abel
type ODEs fully integrated with Maple's ODE-solver {\bf dsolve}.

The classification shown has had the intention of giving a first step
towards organizing in a single reference the integrable cases widely
scattered throughout the literature. The derivation of new solvable
parameterized classes from the works by Abel and Appell in the 19$^{th}$
century (Classes ``C" and ``D") also showed that valuable information can
still be obtained from these old papers. In fact, from Tables 1, 2, 3 and
4 in \se{solvable_seeds}, the larger number of integrable cases found in
textbooks are particular members of this Class ``C" (\eq{class_C}) - an
integrable class derived from a case somehow disregarded in Abel's
Memoires \cite{abel}.

As for the computer routines, the implementation presented here for
solving the equivalence problem for {\it parameterized} classes proved to
be a valuable tool in most of the Abel ODE examples we were able to
collect. The Abel ODE routines here presented were recently integrated
to the Maple computer algebra system and are already part 
of the upcoming Maple release 6.


On the other hand, we note the intrinsic limitation of this Abel ODE
problem: most of the solutions can only be obtained in implicit form and
in terms of quadratures; in turn, these integrals are usually elliptic
integrals so that they cannot be expressed using elementary functions.
Also the classification of integrable cases presented here is incomplete
in that it is missing - at least - a more thorough analysis of the
integrable cases presented in \cite{green}. We are presently working on
this topic and expect to succeed in obtaining reportable results in the
near future.

\ni {\bf Acknowledgments}

\noindent This work was partially supported by the State University of Rio de Janeiro
(UERJ), Brazil and by the Symbolic Computation Group, Faculty of
Mathematics, University of Waterloo, Ontario, Canada. The authors would like
to thank K. von B\"ulow
 for a careful reading of
this paper, and T. Kolokolnikov and A. Wittkopf for fruitful discussions.

\newpage
\appendix
\section*{Appendix A\footnote{The
solution shown for the representative of class D is not valid when $\alpha$
is an integer, or when $2 \alpha$ is a positive integer. In those cases,
the solution of the associated Riccati equation \eq{appell_riccati} takes
many different forms depending on the value of $\alpha$, which we found
inconvenient to present here.
${\rm Ei}(n,x)=\displaystyle {\int}_{\!\!1}^{^{\infty }}{\frac
{{\e^{-xt}}}{{t}^{n}}}{dt}$ is the exponential integral,
${\rm Ai}(x)$ and ${\rm Bi}(x)$ are the Airy wave functions,
${\rm K}(x)$ and ${\rm I}(x)$ are the modified Bessel functions
of the first and second kinds, respectively,
and ${\rm M}(x)$ and ${\rm W}(x)$ are the Whittaker functions.}
}

{\begin{center} {\footnotesize
\begin{tabular}{|c|p{5.8 in}|}
\hline
Class  &  Representative equation and solution

\\
\hline
1 & $
\y1={\fr {3\,y^2-3\,y-x}{x \l(8\,y-9 \r)}},
$
$
\ \ C_1+\fr {x^3\l(4\,x^2+\l(8\,y^2-36\,y+ 27\r)x+4\,y^4-4\,y^3\r)}
{\l(x^2+2\,x \l(y^2-3\,y\r)+y^4 \r)^3} =0
$ \\
\hline
2 & $
\y1=-2\,y^2 x+y^3,
$
$
\ \ C_1+\fr{x{\rm Ai}\l(x^2-\fr{1}{y}\r)+{\rm Ai}\l(1,x^2-\fr{1}{y}\r)}
{x{\rm Bi}\l(x^2-\fr{1}{y}\r)+{\rm Bi}\l(1,x^2-\fr{1}{y}\r)}=0
$ \\
\hline
3 & $
\y1=\frac {y^3}{4\,x^2}-y^2,
$
$
\ \ C_1+\fr{\l(x-\fr{1}{y}\r)
 {\rm Ai}\l(\l(x-\fr{1}{y}\r)^2 -\fr{1}{2x}\r)
+{\rm Ai}\l(1,\l(x-\fr{1}{y}\r)^2-\fr{1}{2x}\r)}
{\l(x-\fr{1}{y}\r)
 {\rm Bi}\l(\l(x-\fr{1}{y}\r)^2-\fr{1}{2x}\r)
+{\rm Bi}\l(1,\l(x-\fr{1}{y}\r)^2-\fr{1}{2x}\r)}=0
$ \\
\hline
4 & $
\y1=y^3-\fr{x+1}{x} y^2,
$
$
\ \ C_1+\fr{1}{x}e^{\fr{1}{y}-x}-{\rm Ei}(1,x-\fr{1}{y})=0
$ \\
\hline
5 & $
\y1=-{\frac {\left (2\,x+3\right )\left (x+1\right ){y}^{3}}
{2\,{x}^{5}}}+{\frac {\left (5\,x+8\right ){y}^{2}}{2\,{x}^{3}}},
$

$
C_1+{\frac {\sqrt {A}}{\sqrt [4]{4\,{\frac {\left (x+1\right )^{2}}
{{x}^{2}\,A}}+1}}}+{\int}^{2\,{\frac {x+1}{x\sqrt {A}}}}\!\left ({z
}^{2}+1\right )^{-5/4}{dz}=0,
\ \ A=\fr{4}{y}-\fr{10}{x}-\fr{6}{x^{2}}-4
$ \\
\hline
6 & $
\y1=-{\frac {{y}^{3}}{{x}^{2}\left (x-1\right )^{2}}}+{\frac {\left (
1-x-{x}^{2}
\right ){y}^{2}}{{x}^{2}\left (x-1\right )^{2}}},
$
$
\ \ C_1-{\rm Ei}\l(1,{\frac {y+{x}^{2}-x}{xy\left (x-1\right )}}\r)
+
\fr
{\left (x-1\right )\,y\,{\e^{{\frac {x-y-{x}^{2}}{xy\left (x-1\right )}}}}}
{x-1+y}=0
$ \\
\hline
7 & $
\y1={\frac {\left (4\,{x}^{4}+5\,{x}^{2}+1\right ){y}^{3}}{2\,{x}^{3
}}}+{y}^{2}
+{\frac {\left (1- 4\,{x}^{2}\right )y}{2\,x\left ({x}^{2}
+1\right )}}
$

$
C_1
+ 2\,{\frac {x+A}{\sqrt [4]{{A}^{2}+1}\left (Ax-1\right )}}
+\int ^{A}\!\left ({z}^{2}+1\right )^{-5/4}{dz}=0
\displaystyle
\ \
A={\frac {x - 2\,y{x}^{4}-3\,y{x}^{2}-y}{x\left (x+y{x}^{2}+y\right )}}
$ \\
\hline
A & $
\y1=\left (\alpha\,x+\fr{1}{x}+\fr{1}{x^{3}}\right ){y}^{3}+{y}^{2},
$

$
C_1+\fr {x^3}{y+x}\,
 \exp \l({\int} ^{\frac {- y x^2}{y+x}}\!
\frac {2\, dz}{z^2-z-\alpha z^3 } \r)
-{\int} ^{\frac {- y x^2}{y+x}}\!
 \exp \l(\int \! \frac {2\, dz}{z^2-z-\alpha z^3 } \r)
{dz}=0
$ \\
\hline
B & $
\y1=2\,\l(x^2-\alpha^2\r)y^3+2\,\l(x+1\r)y^2,
$

$
C_1+\fr
{\l(\alpha+x\r){\rm K}\l(\alpha,-\sqrt{x^2+\fr{1}{y}-\alpha^2}\r)
+\sqrt{x^2+\fr{1}{y}-\alpha^2}\ {\rm K}\l(1+\alpha,-\sqrt{x^2+\fr{1}{y}-\alpha^2}\r)}
{\l(\alpha+x\r){\rm I}\l(\alpha,-\sqrt{x^2+\fr{1}{y}-\alpha^2}\r)
-\sqrt{x^2+\fr{1}{y}-\alpha^2}\ {\rm I}\l(1+\alpha,-\sqrt{x^2+\fr{1}{y}-\alpha^2}\r)}
=0
$ \\
\hline
C & $
\y1={\frac {\alpha\left (1-{x}^{2}\right ){y}^{3}}{2\,x}}+\left (\alpha-1
\right ){y}^{2}-{\frac {\alpha y}{2\,x}},
$
$
\ \ C_1+\fr{\alpha}{x}\left (1-\l(\fr{1}{y}-x\r)^{2}\right )^{{1/\alpha}}
-2\,{\int}^{^{\frac {1-xy}{y}}}\!\left (1-{z}^{2}\right )^{
{\frac {1-\alpha}{\alpha}}}{dz}=0
$ \\
\hline
D & $
\y1=-{\frac {{y}^{3}}{x}}-{\frac {\left (\alpha+{x}^{2}\right ){y}^{2}}{{x}
^{2}}},
$

$
C_1+\fr      {\l(\alpha +1\r){\rm M}\l(-\fr{\alpha}{2}\,
-\fr{3}{4},\fr{1}{4},\fr{1}{2}\,\l(x-\fr{\alpha }{x}-\fr{1}{y}\r)^2\r)
+ \l(\fr{x}{y}-x^2\r){\rm M}\l(-\fr{\alpha}{2}+\fr{1}{4},\fr{1}{4},\fr{1}{2}\,
\l(x-\fr{\alpha }{x}-\fr{1}{y}\r)^2\r)}
   {\l(\alpha^2+\alpha\r){\rm W}\l(-\fr{\alpha}{2}\,
-\fr{3}{4},\fr{1}{4},\fr{1}{2}\,\l(x-\fr{\alpha }{x}-\fr{1}{y}\r)^2\r)
+2\l(\fr{x}{y}-x^2\r){\rm W}\l(-\fr{\alpha}{2}+\fr{1}{4},\fr{1}{4},\fr{1}{2}\,
\l(x-\fr{\alpha }{x}-\fr{1}{y}\r)^2\r)
}=0
$ \\
\hline
\multicolumn{2}{c}
{Representative ODEs and their solutions
 for the Abel ODE classes presented in this work.}
\end{tabular}}
\end{center}}

\label{lastpage}
\end{document}